\documentclass[a4paper,12pt]{article}
\pdfoutput=1 

\usepackage{amsmath,amssymb,bm,bbm,mathtools,amsthm,ascmac}
\usepackage[thicklines]{cancel}
\usepackage{colortbl}
\usepackage{datetime2}
\usepackage[mathscr]{eucal}
\usepackage[bottom]{footmisc}
\usepackage{graphicx} 
\usepackage{indentfirst}
\usepackage{marvosym,wasysym}
\usepackage{multicol}
\usepackage{multirow}
\usepackage[numbers]{natbib} 
\usepackage{tabularx,arydshln}
 \usepackage{ulem}
\usepackage{xcolor} 
\usepackage{cancel}

\definecolor{brown}{HTML}{A52A2A}
\definecolor{darkcyan}{HTML}{008B8B}
\definecolor{dodgerblue}{HTML}{1E90FF}
\definecolor{orangered}{HTML}{FF4500}
\definecolor{royalblue}{HTML}{2B60DE}

\definecolor{refkey}{HTML}{D0A0A0}
\definecolor{labelkey}{HTML}{556B2F}

\usepackage[a4paper,top=36truemm,bottom=36truemm,left=17truemm,right=17truemm,textwidth=176truemm,textheight=230.0truemm]{geometry}

\usepackage[bookmarks=true,bookmarksnumbered=true,bookmarkstype=toc]{hyperref} 
\hypersetup{%
pdftitle={T-duality on Almost Hermitian Spaces},%
pdfauthor={Tetsuji KIMURA; Shin SASAKI; Kenta SHIOZAWA},%
colorlinks={true},%
linkcolor={blue},%
urlcolor={blue},%
filecolor={black},%
citecolor={blue}%
}

\makeatletter

 \@addtoreset{equation}{section}
\makeatother

\makeatletter
\def\tbcaption{\def\@captype{table}\caption}
\def\figcaption{\def\@captype{figure}\caption}
\makeatother

\newcounter{Enumerate}

\DeclareFontFamily{U}{rsf}{}
\DeclareFontShape{U}{rsf}{m}{n}{
  <5> <6> rsfs5 <7> <8> <9> rsfs7 <10-> rsfs10}{}
\DeclareMathAlphabet\Scr{U}{rsf}{m}{n}

\newcommand{\del}{\partial}

\newcommand{\bsubeq}{\begin{subequations}}
\newcommand{\esubeq}{\end{subequations}}

\newcommand{\eps}{\epsilon}
\newcommand{\half}{\frac{1}{2}}
\newcommand{\LS}{\ \ \ \ \ \ \ \ \ \ }
\newcommand{\ls}{\ \ \ \ \ }

\newcommand{\nn}{\nonumber}

\newcommand{\ol}{\overline}

\newcommand{\tpsi}{\tilde{\psi}}

\newcommand{\I}{{\rm i}}
\newcommand{\N}{\mathcal{N}}
\newcommand{\T}{{\rm T}}
\renewcommand{\d}{{\rm d}}
\newcommand{\e}{{\rm e}}

\renewcommand{\r}{\varrho}
\renewcommand{\t}{\vartheta}
\newcommand{\w}{\wedge}
\newcommand{\x}{\text{x}}

\newcommand{\slb}{\scalebox}

\allowdisplaybreaks[4]
\begin{document}
\thispagestyle{empty}
\begin{flushright}
September 2024 
\end{flushright}

\vspace{25mm}

\begin{center}
\slb{1.9}{T-duality on Almost Hermitian Spaces}

\vspace{15mm}

{\large
Tetsuji {\sc Kimura}${}^{*}$, 
Shin {\sc Sasaki}${}^{\dagger}$, 
Kenta {\sc Shiozawa}${}^{\ddagger}$ 
}

\vspace{5mm}

{\small
${}^{*}$
Department of Information and Computer Science, 
Osaka Institute of Technology \\
Hirakata, Osaka 573-0196, JAPAN \\
{\tt tetsuji.kimura \_at\_ oit.ac.jp}
\\
  \vspace{0.3cm}
  ${}^{\dagger}$
  Department of Physics, 
Kitasato University \\
  Sagamihara, Kanagawa 252-0373, JAPAN \\
{\tt shin-s \_at\_ kitasato-u.ac.jp}
\\
  \vspace{0.3cm}
  ${}^{\ddagger}$
  College of Liberal Arts and Sciences,
  Kitasato University \\
  Sagamihara, Kanagawa 252-0373, JAPAN \\
{\tt shiozawa.kenta \_at\_ kitasato-u.ac.jp}
  }
\end{center}

\vspace{10mm}

\begin{abstract}

We investigate T-duality transformation on an almost bi-hermitian space with torsion.
By virtue of the Buscher rule,
we completely describe not only the covariant derivative of geometrical objects but also the Nijenhuis tensor.
We apply this description to 
an almost bi-hermitian space with isometry and investigate integrability on its T-dualized one.
We find that hermiticity is not a sufficient condition to preserve integrability under T-duality transformations.
However, in the presence of the K\"{a}hler condition, 
the T-dualized space still admits integrability of the almost complex structures.
We also observe that the form of H-flux is suitable for string compactification scenarios.

\end{abstract}

\newpage
{\footnotesize
\setlength{\baselineskip}{0mm}
\tableofcontents
}

\section{Introduction}
\label{sect:introduction}

String theory is regarded as one of the most promising candidates which describes classical and quantum dynamics of spacetime geometry.
In particular, string sigma model with worldsheet supersymmetry restricts several strong conditions on its own target space.
It is believed that such conditions are preserved under various string dualities.
S- and U-duality are difficult to analyze spacetime geometry directly because they are non-perturbatively treated.
However, T-duality, defined perturbatively, is suitable for investigation of geometrical features both in the sigma model viewpoint and target space viewpoint.
From the string worldsheet viewpoint, T-duality is quite simple.
This is given by the exchange of the momentum modes and the winding modes.
On the other hand, from the spacetime viewpoint, 
the transformation rule of the geometrical objects is a bit complicated.
This is one of the oldest issues in string duality. 
Simultaneously, however, this is one of the most interesting points to investigate dynamics in string theory, even in the current development.

Under T-duality,
two quantum numbers such as the Kaluza-Klein (KK) momentum number and winding number around a compactified spacetime are exchanged, whilst the string mass spectra are identical.
The Fourier transformation of the KK momentum describes the compactified geometry as gravitational behavior. 
On the other hand, the ``Fourier transformation'' of the winding momentum expresses B-field dynamics on the same compactified spacetime.
From the T-duality aspect, this viewpoint seems asymmetric. 
The winding momentum should also be interpreted as a part of spacetime itself, because the role of the KK and winding momenta are identical under T-duality.
In conventional analyses, 
the B-field $B$, coming from the winding modes, is one of the dynamical fields propagating through spacetime, 
whereas the metric field $g$, coming from the KK modes, describes the dynamics of the spacetime geometry itself.
These two fields and the dilaton field $\phi$ are consolidated into the NS-NS sector of string theory.
In some advanced frameworks, they describe the considering spacetime much more formally.
Generalized geometry (GG) \cite{Hitchin:2003cxu, Gualtieri:2003dx} and double field theory (DFT) \cite{Hull:2009mi, Siegel1993} are two of the most developed theories.
GG and DFT would be quite powerful to understand classical and quantum structure of the spacetime dynamics at a deeper level.
Unfortunately, however, they are still a bit abstract when we go back to a conventional gravity theory with matter fields.
This is partly because that the geometrical feature of the B-field might not be completely utilized, even though the Buscher rule \cite{Buscher1987} has been well established.

In this article, we would like to explore the geometrical feature of the B-field 
through its field-strength (called the H-flux, in common) under T-duality.
We mainly focus on the spacetime geometrical features, rather than the string sigma model itself.
We begin with a geometry with almost complex structure(s) and fundamental two-form(s) as well as metric which is compatible with the almost complex structure(s).
This geometry is referred to as an almost (bi-)hermitian space on which the H-flux would appear as a contorsion tensor.
Imposing certain constraints on the almost (bi-)hermitian space, 
we find many kinds of useful geometries in string theory \cite{Geometry_textbook};
Hermitian, K\"{a}hler, Calabi-Yau, K\"{a}hler with torsion, and so forth.
Analogously, hyper-K\"{a}hler, K3, Taub-NUT, Eguchi-Hanson, hyper-K\"{a}hler with torsion, etc. are realized by some restrictions on an almost (bi-)hypercomplex space.
Performing the T-duality transformation on these spaces,
we discuss universal property of the contorsion tensor (and of the B-field itself) more directly.

Here we enumerate various geometrical interests in string theory.
We perform the T-duality transformation on a space with almost complex structures. 
It is well known that the existence of worldsheet supersymmetry strongly restricts the property of the target space geometry \cite{Gates:1984nk, deWit:1988fk}. 
For instance, invariance of the worldsheet supersymmetry of the string sigma model provides covariant constancy of the almost complex structures on the considering target space.
This constancy is preserved under the T-duality transformation\footnote{T-duality from a rotational Killing symmetry prevents the invariance of worldsheet extended supersymmetry \cite{Bakas:1995hc, Hassan:1995je, Alvarez:1995ai}. \label{rotational_T-dual}}.
It is interesting to understand how the restriction of the target space geometry is relaxed when the worldsheet supersymmetry does not exist in the sigma model.
This investigation will be useful when heterotic or bosonic string theory is considered. 
The target space might be a hermitian non-K\"{a}hler space, for instance, a conformal K\"{a}hler \cite{Ivanov:2000ai} (see \cite{arXiv:2310.13197} for more recent developments).
Since T-duality on heterotic or bosonic string is also applicable, it should be important to explore the T-duality transformed spaces and to analyze whether the integrability condition is preserved or not.
In order for that, we should analyze the T-duality transformation rule of various geometrical objects in a concrete manner.
More generally, we must understand whether integrability of the almost complex structures is preserved or broken under T-duality much more carefully.
As an example, an exotic $5^2_2$-brane background geometry is derived from the Taub-NUT space by T-duality along infinitely arrayed Taub-NUT cores \cite{deBoer:2012ma}.
In this case the ${\rm SU(2)}$ R-symmetry on the background geometry seems to disappear.
We have to investigate existence and integrability of the almost complex structures on this exotic background.

\vspace{3mm}

This article is organized as follows:
In Section \ref{sect:convention},
we write down the conventions in this article.
We first prepare the metric $g_{mn}$, almost complex structure $J^m{}_n$, fundamental two-form $\omega_{mn}$, and their covariant derivatives with respect to the affine connection with a contorsion tensor $C_{mn}{}^p$. 
We also define the Nijenhuis tensor $N^p{}_{mn}$, which determines integrability of the almost complex structures. 
Finally we exhibit the T-duality of $(g, B, \phi)$ and $(J, \omega)$ governed by the Buscher rule. 
In Section \ref{sect:Full},
we write down the full description of various tensors under the T-duality transformation rule. 
They are completely general because we do not impose any restrictions on the string worldsheet or its target space.
In Section \ref{sect:integrability},
we consider the case that the geometry before T-duality is a general bi-hermitian space with isometry while the worldsheet supersymmetry is {\sl not} introduced.
Instead, we assume integrability of the almost complex structures before T-duality.
We find that the integrability condition on a hermitian space without torsion is broken by T-duality\footnote{We note that we only investigate general features of the integrability condition of the almost complex structures on (non-)K\"{a}hler spaces under the T-duality transformation. Analyses of concrete non-K\"{a}hler spaces will be studied in our future works.}.
On the contrary, however, 
the K\"{a}hler condition preserves integrability under the T-duality.
We also find a specific relation on the contorsion tensor on the T-dualized K\"{a}hler space.
Finally, we demonstrate T-duality transformations of the complex projective space $\mathbb{C}P^2$ and Taub-NUT space as the typical K\"{a}hler geometries with isometry.
Section \ref{sect:conclusion} is devoted to the conclusion and discussions.
In Appendix \ref{sect:T-dual_basic},
we exhibit the T-duality transformation of the affine connection derived from the Buscher rule. 
In Appendix \ref{sect:WS_SUSY},
we review integrability of the almost complex structures in the presence of worldsheet supersymmetry.

\section{Conventions}
\label{sect:convention}

We consider an even-dimensional almost bi-hermitian space ${\cal M}$ with isometry.
We always employ adapted local coordinates that make manifest the isometry directions.
The local coordinates on ${\cal M}$ are expressed by $\{x^m \} \supset \{ x^y, x^i \}$.
The indices $m,n,p,q,r,s,\ldots$ run the whole directions on ${\cal M}$.
$y$ goes along the isometry direction,
while $i,j,k,l,h,\ldots$ label the other directions without isometry, respectively.
All geometrical objects are assumed to be independent of the $x^y$-coordinate.
Various basics follow from \cite{Geometry_textbook}.

\subsection{Geometrical objects on almost bi-hermitian space}
\label{sect:AbHS}

First of all, we prepare geometrical objects on ${\cal M}$.
We introduce a metric $g_{mn}$, two commuting almost complex structures $(J_{\pm})^m{}_n$, and two fundamental two-forms $(\omega_{\pm})_{mn}$.
The subscripts $\pm$ are associated with chirality on string worldsheet\footnote{The ``chirality'' of fermions $\theta^{\pm}$ in \cite{Gates:1984nk, deWit:1988fk} links the ``light-cone direction'' $\sigma^{\pm} \coloneqq \tau \pm \sigma$ or ``holomorphy'' \cite{Rocek:1991ps, Witten:1993yc, Hori:2000kt} on two-dimensional string worldsheet coordinates. Thus we use the same subscripts $\pm$ even in bosonic string sigma model in Section \ref{sect:integrability}.}.
The relation among them is described in such a way that 
\bsubeq \label{AHS_property}
\begin{gather}
g_{pq} \, (J_{\pm})^p{}_m \, (J_{\pm})^q{}_n \ = \ g_{mn} 
\, , \quad
(J_{\pm})^m{}_p \, (J_{\pm})^p{}_n \ = \ - \delta^m_n
\, , \\
g_{mn} \ = \ 
(\omega_{\pm})_{mp} \, (J_{\pm})^p{}_n
\, , \quad
(J_{\pm})^m{}_n \ = \ - g^{mp} \, (\omega_{\pm})_{pn}
\, , \quad
(\omega_{\pm})_{mn} \ = \ 
- g_{mp} \, (J_{\pm})^p{}_n
\, , \label{gJomega} \\
(J_+)^m{}_p \, (J_-)^p{}_n - (J_-)^m{}_p \, (J_+)^p{}_n
\ = \ 
0
\, . \label{comm_JJ}
\end{gather}
\esubeq
To investigate geometrical aspects of the space ${\cal M}$,
we introduce the Christoffel symbol $\Gamma_{mn}{}^p$ associated with the metric
\begin{align}
\Gamma_{mn}{}^p
\ &\coloneqq \ 
\half \, g^{pq} \, (\del_m g_{qn} + \del_n g_{mq} - \del_q g_{mn})
\ = \ 
\Gamma_{nm}{}^p
\, . \label{Christoffel_conn} 
\end{align}
By using this, we define covariant derivatives
\begin{align}
\nabla_m A^n
\ &\coloneqq \
\del_m A^n + \Gamma_{mp}{}^n \, A^p
\, , \qquad
\nabla_m A_n
\ \coloneqq \
\del_m A_n - \Gamma_{mn}{}^p \, A_p
\, . \label{covD_1}
\end{align}
In string theory and its low energy effective theories, many tensors provide various aspects of the background geometry.
They often replace the Christoffel symbol in (\ref{covD_1}) with a general affine connection $\hat{\Gamma}_{mn}{}^p$ in such a way that
\bsubeq \label{covD_2}
\begin{align}
\hat{\nabla}_m A^n
\ &= \ 
\del_m A^n + \hat{\Gamma}_{mp}{}^n \, A^p
\, , \\
\hat{\nabla}_m A_n
\ &= \ 
\del_m A_n - \hat{\Gamma}_{mn}{}^p \, A_p
\, . 
\end{align}
\esubeq
We define the antisymmetric part of the affine connection as a torsion tensor $T_{mn}{}^p$
\begin{align}
T_{mn}{}^p
\ &\coloneqq \ 
2 \hat{\Gamma}_{[mn]}{}^p
\ = \ 
\hat{\Gamma}_{mn}{}^p - \hat{\Gamma}_{nm}{}^p
\, .
\end{align}
Alternatively, we describe the affine connection in (\ref{covD_2}) as the sum of the Christoffel symbol and another tensor $C_{mn}{}^p$
\begin{align}
\hat{\Gamma}_{mn}{}^p 
\ &\coloneqq \ 
\Gamma_{mn}{}^p + \xi \, C_{mn}{}^p
\, . \label{affine_conn}
\end{align}
Here we introduce a constant $\xi$ which will be fixed in due course.
In general, $C_{mn}{}^p$ is regarded as a contorsion tensor. 
The contorsion tensor is related with the torsion tensor in the following way:
\begin{align}
T_{mn}{}^p
\ &= \ 
2 \hat{\Gamma}_{[mn]}{}^p
\ = \
\xi \, (C_{mn}{}^p - C_{nm}{}^p)
\, . \label{def_torsion}
\end{align}
In this article, we only consider that $C_{mn}{}^p$ is given by the field-strength of the B-field $B_{mn}$: 
\begin{align}
C_{mn}{}^p 
\ &= \ 
H_{mnq} \, g^{qp}
\, , \quad 
H_{mnp} 
\ \coloneqq \ 
3 \del_{[m} B_{np]}
\, . \label{H-flux}
\end{align}
Then the contorsion tensor in this article is restricted to a totally antisymmetric tensor. 
Substituting (\ref{H-flux}) into (\ref{def_torsion}), the torsion tensor is also given by the H-flux
\begin{align}
T_{mn}{}^p
\ &= \ 
2 \xi \, H_{mnq} \, g^{qp}
\, .
\end{align}

In general relativity, the metricity condition is introduced as 
\begin{align}
\nabla_m g_{np} \ = \ 0
\, .
\label{metricity_cond}
\end{align}
This leads to a set of equations
\begin{gather}
\Gamma_{ym}{}^p \, g_{pn}
\ = \ 
- \Gamma_{yn}{}^p \, g_{pm}
\, , \quad
\Gamma_{yy}{}^m \, g_{my} 
\ = \ 0
\, , \quad
\del_i \Big( \frac{1}{g_{yy}} \Big)
\ = \ 
- \frac{2}{(g_{yy})^2} \, \Gamma_{yy}{}^m \, g_{mi}
\, , \label{gyy_eqs}
\end{gather}
where the $x^y$-direction has isometry, i.e., $\del_y g_{mn} = 0$.
If this set were not involved, the expressions in Section \ref{sect:Full} would become complicated extraordinarily. 
In the presence of contorsion,
the metricity condition (\ref{metricity_cond}) seems to be generalized to 
\begin{align}
\hat{\nabla}_m g_{np} \ = \ 0
\, . \label{metricity_2}
\end{align}
This is indeed identical with (\ref{metricity_cond}) because the metric is symmetric $g_{mn} = g_{nm}$ whereas the contorsion tensor is antisymmetric $C_{mn}{}^p = - C_{nm}{}^p$.
For later convenience, 
it is preferred to describe torsionful covariant derivatives $\hat{\nabla}_m (J_{\pm})^p{}_n$ and $\hat{\nabla}_m (\omega_{\pm})_{np}$ explicitly:
\bsubeq \label{hnabla_Jomega}
\begin{align}
\hat{\nabla}_m (J_{\pm})^p{}_n
\ &= \ 
\del_m (J_{\pm})^p{}_n 
+ \hat{\Gamma}_{mq}{}^p \, (J_{\pm})^q{}_n 
- \hat{\Gamma}_{mn}{}^q \, (J_{\pm})^p{}_q 
\nn \\
\ &= \ 
\nabla_m (J_{\pm})^p{}_n
+ \xi_{\pm} \, \big\{ C_{mq}{}^p \, (J_{\pm})^q{}_n - C_{mn}{}^q \, (J_{\pm})^p{}_q \big\}
\, , \label{hnabla_J} \\
\hat{\nabla}_m (\omega_{\pm})_{np}
\ &= \ 
\del_m (\omega_{\pm})_{np}
- \hat{\Gamma}_{mn}{}^q \, (\omega_{\pm})_{qp}
- \hat{\Gamma}_{mp}{}^q \, (\omega_{\pm})_{nq}
\nn \\
\ &= \ 
\nabla_m (\omega_{\pm})_{np}
- \xi_{\pm} \, \big\{ C_{mn}{}^q \, (\omega_{\pm})_{qp} + C_{mp}{}^q \, (\omega_{\pm})_{nq} \big\}
\, . \label{hnabla_omega}
\end{align}
\esubeq
Notice that the constant $\xi$ in (\ref{affine_conn}) is replaced with $\xi_{\pm}$ whose subscripts represent the worldsheet chirality (or holomorphy). 
We will fix them when the Buscher rule \cite{Buscher1987} is discussed.

It is now ready to define the Nijenhuis tensor of the respective $J_{\pm}$ in such a way that
\bsubeq \label{Nijenhuis}
\begin{align}
(N_{\pm})^p{}_{mn}
\ \coloneqq& \ \
(J_{\pm})^q{}_m \big\{ \del_q (J_{\pm})^p{}_n - \del_n (J_{\pm})^p{}_q \big\}
- (J_{\pm})^q{}_n \big\{ \del_q (J_{\pm})^p{}_m - \del_m (J_{\pm})^p{}_q \big\}
\nn \\
\ =& \ \
- (N_{\pm})^p{}_{nm}
\, . \label{N}
\end{align}
Introducing the affine connection (\ref{affine_conn}), 
we generalize this form to 
\begin{align}
(\hat{N}_{\pm})^p{}_{mn}
\ \coloneqq& \ \
(J_{\pm})^q{}_m \big\{ \hat{\nabla}_q (J_{\pm})^p{}_n - \hat{\nabla}_n (J_{\pm})^p{}_q \big\}
- (J_{\pm})^q{}_n \big\{ \hat{\nabla}_q (J_{\pm})^p{}_m - \hat{\nabla}_m (J_{\pm})^p{}_q \big\}
\nn \\
\ =& \ \
- (\hat{N}_{\pm})^p{}_{nm}
\, . \label{Nhat}
\end{align}
The difference between (\ref{N}) and (\ref{Nhat}) is nontrivial 
and will play an important role in analyzing geometrical aspects under T-duality transformations.
It is worth introducing a new terminology $(\Scr{C}_{\pm})^p{}_{mn}$, referred to as $\Scr{C}$-tensors in this article\footnote{Originally, this comes from the $X$-tensors in \cite{deWit:1988fk}. $X^{\pm}_{mnp}$ and $(\Scr{C}_{\pm})^p{}_{mn}$ are related with $X^{\pm}_{mnp} = (\Scr{C}_{\pm})_{mnq} \, (J_{\pm})^q{}_p$.}:
\begin{align}
(\hat{N}_{\pm})^p{}_{mn} - (N_{\pm})^p{}_{mn}
\ &= \ 
2 \, \xi_{\pm} \, g^{pq} \, \Big\{
3 \, (J_{\pm})^r{}_{[m} \, (J_{\pm})^s{}_{n} \, H_{q]rs}
- H_{mnq}
\Big\}
\nn \\
\ &\eqqcolon \
2 \, \xi_{\pm} \, (\Scr{C}_{\pm})^p{}_{mn}
\, . \label{ScrC_def}
\end{align}
\esubeq

\subsection{T-duality transformation rule}
\label{sect:T-dual_trsf}

Here we exhibit the T-duality transformation rule of the geometrical objects $(g, B, \phi)$ and $(J_{\pm}, \omega_{\pm})$ on an almost bi-hermitian space ${\cal M}$ to $(g', B', \phi')$ and $(J'_{\pm}, \omega'_{\pm})$ on a T-dualized space ${\cal M}'$. 
We perform the T-duality transformation along the $x^y$-direction which has isometry.
The rule of the metric $g_{mn}$, B-field $B_{mn}$ and dilaton $\phi$ is represented by Buscher \cite{Buscher1987}\footnote{In this article we only consider the case that $\Scr{L}_K B = 0$, 
where $\Scr{L}_K$ is the Lie derivative along the Killing direction $K$. 
It is also noticed that the existence of an isometry in the original space ${\cal M}$ is not sufficient for T-duality if there are obstructions to gauging the isometry. 
Detailed discussions are given in \cite{Hull:2006qs, Abou-Zeid:2019lgj}, 
where the Buscher rule (\ref{Buscher-rule}) is generalized if $\Scr{L}_K B \neq 0$.}:
\bsubeq \label{Buscher-rule}
\begin{alignat}{3}
g'_{ij}
\ &= \ 
g_{ij} - \frac{g_{iy} \, g_{jy} - B_{iy} \, B_{jy}}{g_{yy}} 
\, , &\quad
g'_{iy}
\ &= \
\frac{B_{iy}}{g_{yy}} \ = \ g'_{yi} 
\, , &\quad
g'_{yy}
\ &= \ 
\frac{1}{g_{yy}}
\, , \\
B'_{ij}
\ &= \ 
B_{ij} - \frac{B_{iy} \, g_{jy} - g_{iy} \, B_{jy}}{g_{yy}}
\, , &\quad
B'_{iy} 
\ &= \ 
\frac{g_{iy}}{g_{yy}} \ = \ - B'_{yi}
\, , &\quad
\phi'
\ &= \ 
\phi - \half \log (g_{yy})
\, , 
\end{alignat}
where we assume that all of the fields are independent of the coordinate $x^y$. 
We notice that the above expression is obtained by setting the constants $\xi_{\pm}$ to 
\begin{align}
\xi_{\pm} \ &\equiv \ \pm \half 
\, . \label{constants_fixed}
\end{align}
It is worth describing the Buscher rule of the inverse metric $g^{mn}$ as follows:
\begin{align}
g'{}^{ij}
\ &= \ 
g^{ij}
\, , \quad
g'{}^{iy}
\ = \ 
B_{ym} \, g^{mi}
\, , \quad
g'{}^{yy}
\ = \ 
- B_{ym} \, g^{mn} \, B_{ny} + g_{yy}
\, . 
\end{align}
\esubeq
The T-duality transformation rule of $(J_{\pm})^m{}_n$ and $(\omega_{\pm})_{mn}$ is acquired \cite{Hassan:1994mq, Kimura:2022dma}\footnote{Although the rule (\ref{KSS-rule}) was first derived in the context of supersymmetric string sigma models, we emphasize that they remain valid even in non-supersymmetric cases \cite{Hull:2009mi}.}
as
\bsubeq \label{KSS-rule}
\begin{align}
(J'_{\pm})^i{}_j
\ &= \ 
(J_{\pm})^i{}_j
- \frac{(J_{\pm})^i{}_y \, (g_{jy} \pm B_{jy})}{g_{yy}}
\, , \\
(J'_{\pm})^i{}_y
\ &= \ 
\mp \frac{(J_{\pm})^i{}_y}{g_{yy}}
\, , \\
(J'_{\pm})^y{}_j
\ &= \ 
\pm (\omega_{\pm})_{yj}
+ B_{ym} \, \Big\{
(J_{\pm})^m{}_j
- \frac{(J_{\pm})^m{}_y \, (g_{jy} \pm B_{jy})}{g_{yy}}
\Big\}
\, , \\
(J'_{\pm})^y{}_y
\ &= \ 
\mp \frac{B_{ym} \, (J_{\pm})^m{}_y}{g_{yy}}
\, , \\
(\omega'_{\pm})_{ij}
\ &= \ 
(\omega_{\pm})_{ij}
- \frac{(\omega_{\pm})_{iy} \, (g_{jy} \pm B_{jy}) - (g_{iy} \pm B_{iy}) \, (\omega_{\pm})_{jy}}{g_{yy}}
\, , \\
(\omega'_{\pm})_{iy}
\ &= \ 
\mp \frac{(\omega_{\pm})_{iy}}{g_{yy}}
\, .
\end{align}
\esubeq
Under the above transformation rule,
the property of almost bi-hermitian space (\ref{AHS_property}) is preserved 
\bsubeq \label{gJomega_T}  
\begin{gather}
g'_{pq} \, (J'_{\pm})^p{}_m \, (J'_{\pm})^q{}_n \ = \ g'_{mn}
\, , \quad
(J'_{\pm})^m{}_p \, (J'_{\pm})^p{}_n 
\ = \ 
- \delta^m_n
\, , \\
g'_{mn} \ = \ 
(\omega'_{\pm})_{mp} \, (J'_{\pm})^p{}_n
\, , \quad
(J'_{\pm})^m{}_n \ = \ - g'{}^{mp} \, (\omega'_{\pm})_{pn}
\, , \quad
(\omega'_{\pm})_{mn} \ = \ 
- g'_{mp} \, (J'_{\pm})^p{}_n
\, , \\
(J'_+)^m{}_p \, (J'_-)^p{}_n - (J'_-)^m{}_p \, (J'_+)^p{}_n
\ = \ 
0
\, . \label{comm_JJ_T}
\end{gather}
\esubeq

\subsection{Almost bi-hypercomplex space}
\label{sect:bi-hypercomplex}

The property of almost bi-hermitian space (\ref{AHS_property}) is applied to that of almost bi-hypercomplex space in which three sets of almost complex structures $J_{a,\pm}$ (where the subscript runs $a = 1,2,3$) are involved:
\bsubeq \label{gJomega_a}  
\begin{gather}
g_{pq} \, (J_{a,\pm})^p{}_m \, (J_{a,\pm})^q{}_n \ = \ g_{mn}
\, , \quad
(J_{a,\pm})^m{}_p \, (J_{b,\pm})^p{}_n
\ = \ 
- \delta_{ab} \, \delta^m_n 
+ \eps_{abc} \, (J_{c,\pm})^m{}_n
\, , \\
g_{mn} \ = \ (\omega_{a,\pm})_{mp} \, (J_{a,\pm})^p{}_n
\, , \quad
(J_{a,\pm})^m{}_n \ = \ - g{}^{mp} \, (\omega_{a,\pm})_{pn}
\, , \quad 
(\omega_{a,\pm})_{mn} \ = \ - g_{mp} \, (J_{a,\pm})^p{}_n
\, , \\
(J_{a,+})^m{}_p \, (J_{b,-})^p{}_n - (J_{b,-})^m{}_p \, (J_{a,+})^p{}_n
\ = \ 0
\, . \label{comm_JJ_a}
\end{gather}
\esubeq
$\eps_{abc}$ is the Levi-Civita totally antisymmetric symbol with normalization $\eps_{123} = +1$.
Analogously, the tensors 
$(N_{a,\pm})^p{}_{mn}$, 
$(\hat{N}_{a,\pm})^p{}_{mn}$, and
$(\Scr{C}_{a,\pm})^p{}_{mn}$
are constructed as in (\ref{Nijenhuis}).
The T-duality transformation rule is completely parallel to (\ref{Buscher-rule}) and (\ref{KSS-rule}).
For detailed discussions, see \cite{Kimura:2022dma}.

\section{Full description}
\label{sect:Full}

In this section, we describe various tensors after T-duality transformation\footnote{The affine connection, which is not a tensor, is also transformed by T-duality. They are summarized in Appendix \ref{sect:T-dual_basic}. 
They are the basic objects to investigate general covariance in the presence of torsion.}. 
Our purpose is to evaluate integrability of the almost complex structures on ${\cal M}'$.
Therefore, as a preparation, we perform the T-duality transformation of the covariant derivatives of the metric, almost complex structures, and fundamental two-forms. 
After that, we evaluate the T-dualized expression of the Nijenhuis tensors and $\Scr{C}$-tensors.
We notice that we do not impose any conditions 
beyond (\ref{AHS_property}) and (\ref{metricity_2}) on the space ${\cal M}$.
Hence we expect that the space ${\cal M}'$ would also be an almost bi-hermitian space with commuting almost complex structures (\ref{gJomega_T}) without any further restrictions.

\subsection{Metricity condition}
\label{sect:metricity_T}

First of all, we should discuss the metric $g'_{mn}$ satisfies the metricity condition (\ref{metricity_2}) after the T-duality transformation.
This is observed in terms of the T-duality transformation of the affine connection in Appendix \ref{sect:T-dual_basic} in such a way that\footnote{We should notice that $\del'_m \coloneqq \del / \del x'{}^m$ is a derivative on ${\cal M}'$. 
Now T-duality is performed along the $x^y$-direction, rather than $x^i$-directions; i.e., $x'{}^i = x^i$. 
In Section \ref{sect:convention}, we have assumed that any function $f'(x')$ is independent of $x'{}^y$ (and $x^y$). 
Thus we observe $\del'_i f'(x')= \del_i f'(x')$, $\del'_y f'(x') = 0 = \del_y f'(x')$, and $f'(x') = f'(x)$.
Consequently $\del'_m f'(x')$ becomes identical with $\del_m f'(x)$.}
\begin{align}
\hat{\nabla}'_m g'_{np} \ = \ 0
\, . \label{metricity_3}
\end{align}
By virtue of this result, we can proceed other computations of the T-duality transformation in the presence of torsionful covariant derivatives.

\subsection{Covariant derivative of almost complex structures}
\label{sect:covD_J_T}

Let us perform the T-duality transformation of the covariant derivatives of the almost complex structures $\hat{\nabla}_m (J_{\pm})^p{}_n$ (\ref{hnabla_J}) along the $x^y$-direction.
They will play a central role in string sigma model in the presence of worldsheet supersymmetry \cite{Gates:1984nk, deWit:1988fk}.
The T-dualized values $\hat{\nabla}'_m (J'_{\pm})^p{}_n$ are represented by the original ones:
\bsubeq \label{hnabla_J_T_}
\begin{align}
\hat{\nabla}'_y (J'_{\pm})^y{}_y
\ &= \ 
- \frac{1}{(g_{yy})^2} \, B_{ym} \, \hat{\nabla}_y (J_{\pm})^m{}_y 
\, , \\
\hat{\nabla}'_i (J'_{\pm})^y{}_y
\ &= \ 
\mp \frac{1}{g_{yy}} \, B_{ym} \, \hat{\nabla}_i (J_{\pm})^m{}_y
\pm \frac{1}{(g_{yy})^2} \, (g_{iy} \mp B_{iy}) \, B_{ym} \, \hat{\nabla}_y (J_{\pm})^m{}_y
\, , \\
\hat{\nabla}'_y (J'_{\pm})^y{}_k
\ &= \ 
- \frac{1}{g_{yy}} \, (g_{ym} \mp B_{ym}) \, \hat{\nabla}_y (J_{\pm})^m{}_k
\mp \frac{1}{(g_{yy})^2} \, (g_{yk} \mp B_{yk}) \, B_{ym} \, \hat{\nabla}_y (J_{\pm})^m{}_y 
\, , \\
\hat{\nabla}'_i (J'_{\pm})^y{}_k
\ &= \ 
\mp (g_{ym} \mp B_{ym}) \, \hat{\nabla}_i (J_{\pm})^m{}_k
- \frac{1}{g_{yy}} \, (g_{ky} \pm B_{ky}) \, B_{ym} \, \hat{\nabla}_i (J_{\pm})^m{}_y
\nn \\
\ & \ \ \ \
\mp \frac{1}{g_{yy}} \, (g_{iy} \mp B_{iy}) \, \hat{\nabla}_y (\omega_{\pm})_{yk}
- \frac{1}{g_{yy}} \, (g_{iy} \mp B_{iy}) \, B_{ym} \, \hat{\nabla}_y (J_{\pm})^m{}_k
\nn \\
\ & \ \ \ \
+ \frac{1}{(g_{yy})^2} \, (g_{iy} \mp B_{iy}) \, (g_{ky} \pm B_{ky}) \, B_{ym} \, \hat{\nabla}_y (J_{\pm})^m{}_y 
\, , \\
\hat{\nabla}'_y (J'_{\pm})^j{}_y
\ &= \ 
- \frac{1}{(g_{yy})^2} \, \hat{\nabla}_y (J_{\pm})^j{}_y
\, , \\
\hat{\nabla}'_i (J'_{\pm})^j{}_y
\ &= \ 
\mp \frac{1}{g_{yy}} \, \hat{\nabla}_i (J_{\pm})^j{}_y
\pm \frac{1}{(g_{yy})^2} \, (g_{iy} \mp B_{iy}) \, \hat{\nabla}_y (J_{\pm})^j{}_y
\, , \\
\hat{\nabla}'_y (J'_{\pm})^j{}_k
\ &= \ 
\pm \frac{1}{g_{yy}} \, \hat{\nabla}_y (J_{\pm})^j{}_k
\mp \frac{1}{(g_{yy})^2} \, (g_{yk} \mp B_{yk}) \, \hat{\nabla}_y (J_{\pm})^j{}_y 
\, , \\
\hat{\nabla}'_i (J'_{\pm})^j{}_k
\ &= \ 
\hat{\nabla}_i (J_{\pm})^j{}_k
- \frac{1}{g_{yy}} \, (g_{iy} \mp B_{iy}) \, \hat{\nabla}_y (J_{\pm})^j{}_k
- \frac{1}{g_{yy}} \, (g_{ky} \pm B_{ky}) \, \hat{\nabla}_i (J_{\pm})^j{}_y 
\nn \\
\ & \ \ \ \
+ \frac{1}{(g_{yy})^2} \, (g_{iy} \mp B_{iy}) \, (g_{ky} \pm B_{ky}) \, \hat{\nabla}_y (J_{\pm})^j{}_y 
\, . 
\end{align}
\esubeq
We emphasize that the result (\ref{hnabla_J_T_}) is the most general description of the T-dualized covariant derivatives of the almost complex structures. 
In later sections we will introduce appropriate conditions in order to find significant property on the T-dualized space ${\cal M}'$.

\subsection{Covariant derivative of fundamental two-forms}
\label{sect:covD_omega_T}

We perform the T-duality transformation of the covariant derivative of the fundamental two-forms $\hat{\nabla}_m (\omega_{\pm})_{np}$ (\ref{hnabla_omega}).
The result is obtained by the Buscher rule (\ref{Buscher-rule}) and (\ref{KSS-rule}) applied directly to $\hat{\nabla}_m (\omega_{\pm})_{np}$:
\bsubeq \label{hnabla_omega_T_}
\begin{align}
\hat{\nabla}'_y (\omega'_{\pm})_{jy}
\ &= \ 
- \frac{1}{(g_{yy})^2} \, \hat{\nabla}_y (\omega_{\pm})_{jy}
\, , \\
\hat{\nabla}'_i (\omega'_{\pm})_{jy}
\ &= \ 
\mp \frac{1}{g_{yy}} \, \hat{\nabla}_i (\omega_{\pm})_{jy}
\pm \frac{1}{(g_{yy})^2} \, \big( g_{iy} \mp B_{iy} \big) \, \hat{\nabla}_y (\omega_{\pm})_{jy}
\, , \\
\hat{\nabla}'_y (\omega'_{\pm})_{jk}
\ &= \ 
\pm \frac{1}{g_{yy}} \, \hat{\nabla}_y (\omega_{\pm})_{jk}
\pm \frac{1}{(g_{yy})^2} \, (g_{ky} \pm B_{ky}) \, \hat{\nabla}_y (\omega_{\pm})_{yj}
\mp \frac{1}{(g_{yy})^2} \, (g_{jy} \pm B_{jy}) \, \hat{\nabla}_y (\omega_{\pm})_{yk}
\, , \\
\hat{\nabla}'_i (\omega'_{\pm})_{jk}
\ &= \ 
\hat{\nabla}_i (\omega_{\pm})_{jk}
\nn \\
\ & \ \ \ \
- \frac{1}{g_{yy}} \, (g_{iy} \mp B_{iy}) \, \hat{\nabla}_y (\omega_{\pm})_{jk} 
- \frac{1}{g_{yy}} \, (g_{jy} \pm B_{jy}) \, \hat{\nabla}_i (\omega_{\pm})_{yk} 
- \frac{1}{g_{yy}} \, (g_{ky} \pm B_{ky}) \, \hat{\nabla}_i (\omega_{\pm})_{jy} 
\nn \\
\ & \ \ \ \ 
+ \frac{1}{(g_{yy})^2} \, (g_{iy} \mp B_{iy}) \, (g_{ky} \pm B_{ky}) \, \hat{\nabla}_y (\omega_{\pm})_{jy} 
- \frac{1}{(g_{yy})^2} \, (g_{iy} \mp B_{iy}) \, (g_{jy} \pm B_{jy}) \, \hat{\nabla}_y (\omega_{\pm})_{ky} 
\, .
\end{align}
\esubeq
We can check the above expression by substituting the metricity conditions (\ref{metricity_2}) and (\ref{metricity_3}) into the result (\ref{hnabla_J_T_}).
Inversely, of course, we can reconstruct (\ref{hnabla_J_T_}) from (\ref{hnabla_omega_T_}).

\subsection{Nijenhuis tensors}
\label{sect:N_T}

We write down the T-dualized expression of the Nijenhuis tensors $(N_{\pm})^p{}_{mn}$ in (\ref{N}). 
Here we find that each component of $(N'_{\pm})^p{}_{mn}$ is divided into two parts: the terms given by the original Nijenhuis tensors and those by the covariant derivatives:
\bsubeq \label{N'_}
\begin{align}
(N'_{\pm})^y{}_{jy}
\ &= \ 
\mp \frac{1}{g_{yy}} \, B_{ym} \, (N_{\pm})^m{}_{jy}
+ \frac{3}{g_{yy}} \, \hat{\nabla}_{[y} (\omega_{\pm})_{jm]} \, (J_{\pm})^m{}_y
\mp \frac{2}{(g_{yy})^2} \, B_{ym} \, \hat{\nabla}_y (\omega_{\pm})_{yj} \, (J_{\pm})^m{}_y
\, , \\
(N'_{\pm})^i{}_{jy}
\ &= \ 
\mp \frac{1}{g_{yy}} \, (N_{\pm})^i{}_{jy}
\mp \frac{2}{(g_{yy})^2} \, \hat{\nabla}_y (\omega_{\pm})_{yj} \, (J_{\pm})^i{}_y
\, , \\
(N'_{\pm})^y{}_{jk}
\ &= \ 
\mp (g_{ym} \mp B_{ym}) \, (N_{\pm})^m{}_{jk}
+ \frac{1}{g_{yy}} \, B_{ym} \, (N_{\pm})^m{}_{yj} \, (g_{yk} \mp B_{yk}) 
- \frac{1}{g_{yy}} \, B_{ym} \, (N_{\pm})^m{}_{yk} \, (g_{yj} \mp B_{yj}) 
\nn \\
\ & \ \ \ \ 
\mp \hat{\nabla}_y (\omega_{\pm})_{jm} \, (J_{\pm})^m{}_k 
\pm \hat{\nabla}_y (\omega_{\pm})_{km} \, (J_{\pm})^m{}_j 
\nn \\
\ & \ \ \ \ 
\pm \frac{3}{g_{yy}} \, \hat{\nabla}_{[y} (\omega_{\pm})_{jm]} \, (J_{\pm})^m{}_y \, (g_{yk} \mp B_{yk})
\mp \frac{3}{g_{yy}} \, \hat{\nabla}_{[y} (\omega_{\pm})_{km]} \, (J_{\pm})^m{}_y \, (g_{yj} \mp B_{yj})
\nn \\
\ & \ \ \ \
- \frac{2}{g_{yy}} \, \hat{\nabla}_y (\omega_{\pm})_{jk} \, B_{yp} \, (J_{\pm})^p{}_y
+ \frac{2}{(g_{yy})^2} \, \hat{\nabla}_y (\omega_{\pm})_{jy} \, B_{yp} \, (J_{\pm})^p{}_y \, (g_{yk} \mp B_{yk}) 
\nn \\
\ & \ \ \ \  
- \frac{2}{(g_{yy})^2} \, \hat{\nabla}_y (\omega_{\pm})_{ky} \, B_{yp} \, (J_{\pm})^p{}_y \, (g_{yj} \mp B_{yj})
\, , \\
(N'_{\pm})^i{}_{jk}
\ &= \ 
(N_{\pm})^i{}_{jk}
- \frac{1}{g_{yy}} \, (g_{jy} \pm B_{jy}) \, (N_{\pm})^i{}_{yk} 
+ \frac{1}{g_{yy}} \, (g_{ky} \pm B_{ky}) \, (N_{\pm})^i{}_{yj} 
- \frac{2}{g_{yy}} \, \hat{\nabla}_y (\omega_{\pm})_{jk} \, (J_{\pm})^i{}_y 
\nn \\
\ & \ \ \ \
- \frac{2}{(g_{yy})^2} \, (g_{jy} \pm B_{jy}) \, \hat{\nabla}_y (\omega_{\pm})_{ky} \, (J_{\pm})^i{}_y 
+ \frac{2}{(g_{yy})^2} \, (g_{ky} \pm B_{ky}) \, \hat{\nabla}_y (\omega_{\pm})_{jy} \, (J_{\pm})^i{}_y 
\, .
\end{align}
\esubeq

\subsection{$\Scr{C}$-tensors}
\label{sect:ScrC_T}

Finally we write down the T-dualized $\Scr{C}$-tensors $(\Scr{C}_{\pm})^p{}_{mn}$ in (\ref{ScrC_def}).
Analogous to the Nijenhuis tensors $(N'_{\pm})^p{}_{mn}$ (\ref{N'_}), the expression $(\Scr{C}'_{\pm})^p{}_{mn}$ is divided into two parts:
the terms given by the original $\Scr{C}$-tensors and those by the covariant derivatives:
\bsubeq \label{ScrC'-2}
\begin{align}
(\Scr{C}'_{\pm})^y{}_{jy}
\ &= \ 
\mp \frac{1}{g_{yy}} \, B_{ym} \, (\Scr{C}_{\pm})^m{}_{jy}
\nn \\
\ & \ \ \ \
+ \hat{\nabla}_y (\omega_{\pm})_{jy} \, \Big\{ + \frac{2}{(g_{yy})^2} \, (J_{\pm})^m{}_y \, B_{my} \Big\}
+ \hat{\nabla}_y (J_{\pm})^m{}_y \, \Big\{ + \frac{2}{(g_{yy})^2} \, (\omega_{\pm})_{yj} \, B_{my} \Big\}
\nn \\
\ & \ \ \ \
+ \hat{\nabla}_y (J_{\pm})^m{}_j \, \Big\{ - \frac{2}{g_{yy}} \, (J_{\pm})^n{}_m \, B_{ny} \Big\}
+ \hat{\nabla}_y (J_{\pm})^m{}_y \, \Big\{ + \frac{2}{(g_{yy})^2} \, (J_{\pm})^n{}_m \, B_{ny} \, g_{jy} \Big\}
\, , \\
(\Scr{C}'_{\pm})^i{}_{jy}
\ &= \ 
\mp \frac{1}{g_{yy}} \, (\Scr{C}_{\pm})^i{}_{jy} 
\nn \\
\ & \ \ \ \
+ \hat{\nabla}_y (J_{\pm})^m{}_j \, \Big\{ + \frac{2}{g_{yy}} \, (J_{\pm})^i{}_m \Big\}
+ \hat{\nabla}_y (J_{\pm})^m{}_y \, \Big\{ - \frac{2}{(g_{yy})^2} \, (J_{\pm})^i{}_m \, g_{jy} \Big\}
\nn \\
\ & \ \ \ \
+ \hat{\nabla}_y (J_{\pm})^i{}_y \, \Big\{ + \frac{2}{(g_{yy})^2} \, (\omega_{\pm})_{jy} \Big\} 
+ \hat{\nabla}_y (\omega_{\pm})_{jy} \, \Big\{ - \frac{2}{(g_{yy})^2} \, (J_{\pm})^i{}_y \Big\} 
\, , \\
(\Scr{C}'_{\pm})^y{}_{jk}
\ &= \ 
\mp (g_{my} \pm B_{my}) \, (\Scr{C}_{\pm})^m{}_{jk} 
- \frac{1}{g_{yy}} \, B_{ym} \, (\Scr{C}_{\pm})^m{}_{yk} \, (g_{jy} \pm B_{jy}) 
- \frac{1}{g_{yy}} \, B_{ym} \, (\Scr{C}_{\pm})^m{}_{jy} \, (g_{ky} \pm B_{ky})
\nn \\
\ & \ \ \ \ 
+ \hat{\nabla}_y (J_{\pm})^m{}_j \, 
\Big\{
- \frac{2}{g_{yy}} \, (J_{\pm})^n{}_m \, B_{ny} \, B_{ky}
\mp \frac{2}{g_{yy}} \, (\omega_{\pm})_{my} \, B_{ky} 
\mp \frac{2}{g_{yy}} \, (\omega_{\pm})_{ky} \, B_{my} 
\Big\} 
\nn \\
\ & \ \ \ \ 
+ \hat{\nabla}_y (J_{\pm})^m{}_k \, 
\Big\{
+ \frac{2}{g_{yy}} \, (J_{\pm})^n{}_m \, B_{ny} \, B_{jy} 
\pm \frac{2}{g_{yy}} \, (\omega_{\pm})_{my} \, B_{jy} 
\pm \frac{2}{g_{yy}} \, (\omega_{\pm})_{jy} \, B_{my} 
\Big\} 
\nn \\
\ & \ \ \ \ 
+ \hat{\nabla}_y (J_{\pm})^m{}_y \, 
\Big\{
- \frac{2}{g_{yy}} \, (\omega_{\pm})_{mj} \, (g_{ky} \pm B_{ky}) 
+ \frac{2}{g_{yy}} \, (\omega_{\pm})_{mk} \, (g_{jy} \pm B_{jy})
\nn \\
\ &\LS\LS\ls
+ \frac{2}{(g_{yy})^2} \, (J_{\pm})^n{}_m \, B_{ny} \, B_{ky} \, g_{jy}
- \frac{2}{(g_{yy})^2} \, (J_{\pm})^n{}_m \, B_{ny} \, B_{jy} \, g_{ky}
\nn \\
\ &\LS\LS\ls
\pm \frac{2}{(g_{yy})^2} \, (\omega_{\pm})_{my} \, B_{ky} \, g_{jy} 
\mp \frac{2}{(g_{yy})^2} \, (\omega_{\pm})_{my} \, B_{jy} \, g_{ky} 
\nn \\
\ &\LS\LS\ls
\mp \frac{2}{(g_{yy})^2} \, (\omega_{\pm})_{yk} \, B_{my} \, (g_{jy} \pm B_{jy}) 
\pm \frac{2}{(g_{yy})^2} \, (\omega_{\pm})_{yj} \, B_{my} \, (g_{ky} \pm B_{ky}) 
\Big\} 
\nn \\
\ & \ \ \ \ 
+ \hat{\nabla}_y (\omega_{\pm})_{mj} \, \Big\{ - (J_{\pm})^m{}_k \Big\} 
+ \hat{\nabla}_y (\omega_{\pm})_{mk} \, \Big\{ + (J_{\pm})^m{}_j \Big\} 
\nn \\
\ & \ \ \ \ 
+ \hat{\nabla}_y (\omega_{\pm})_{jk} \, \Big\{ \mp \frac{2}{g_{yy}} \, (J_{\pm})^m{}_y \, B_{my} \Big\}
\nn \\
\ & \ \ \ \ 
+ \hat{\nabla}_y (\omega_{\pm})_{ky} \, 
\Big\{
- \frac{2}{g_{yy}} \, (\omega_{\pm})_{yj} 
\mp \frac{2}{(g_{yy})^2} \, (J_{\pm})^m{}_y \, B_{my} \, (g_{jy} \pm B_{jy})
\Big\} 
\nn \\
\ & \ \ \ \ 
+ \hat{\nabla}_y (\omega_{\pm})_{jy} \, 
\Big\{
+ \frac{2}{g_{yy}} \, (\omega_{\pm})_{yk}
\pm \frac{2}{(g_{yy})^2} \, (J_{\pm})^m{}_y \, B_{my} \, (g_{ky} \pm B_{ky})
\Big\} 
\, , \\
(\Scr{C}'_{\pm})^i{}_{jk}
\ &= \ 
(\Scr{C}_{\pm})^i{}_{jk}
- \frac{1}{g_{yy}} \, (\Scr{C}_{\pm})^i{}_{jy} \, (g_{ky} \pm B_{ky})
- \frac{1}{g_{yy}} \, (\Scr{C}_{\pm})^i{}_{yk} \, (g_{jy} \pm B_{jy}) 
\nn \\
\ & \ \ \ \
+ \hat{\nabla}_y (J_{\pm})^i{}_j \, 
\Big\{
\mp \frac{2}{g_{yy}} \, (\omega_{\pm})_{yk}
\Big\} 
+ \hat{\nabla}_y (J_{\pm})^i{}_k \, 
\Big\{
\pm \frac{2}{g_{yy}} \, (\omega_{\pm})_{yj}
\Big\} 
\nn \\
\ & \ \ \ \ 
+ \hat{\nabla}_y (J_{\pm})^i{}_m \, 
\Big\{
- \frac{2}{g_{yy}} \, (J_{\pm})^m{}_j \, B_{ky} 
+ \frac{2}{g_{yy}} \, (J_{\pm})^m{}_k \, B_{jy} 
\nn \\
\ &\LS\LS\ls
- \frac{2}{(g_{yy})^2} \, (J_{\pm})^m{}_y \, B_{jy} \, g_{ky}
+ \frac{2}{(g_{yy})^2} \, (J_{\pm})^m{}_y \, B_{ky} \, g_{jy}
\Big\} 
\nn \\
\ & \ \ \ \ 
+ \hat{\nabla}_y (\omega_{\pm})_{jk} \, 
\Big\{
\pm \frac{2}{g_{yy}} \, (J_{\pm})^i{}_y
\Big\} 
\nn \\
\ & \ \ \ \ 
+ \hat{\nabla}_y (\omega_{\pm})_{jy} \, 
\Big\{
\mp \frac{2}{(g_{yy})^2} \, (J_{\pm})^i{}_y \, (g_{ky} \pm B_{ky})
\Big\} 
+ \hat{\nabla}_y (\omega_{\pm})_{yk} \, 
\Big\{
\mp \frac{2}{(g_{yy})^2} \, (J_{\pm})^i{}_y \, (g_{jy} \pm B_{jy})
\Big\} 
\nn \\
\ & \ \ \ \
+ \hat{\nabla}_y (J_{\pm})^i{}_y \, 
\Big\{
\mp \frac{2}{(g_{yy})^2} \, (\omega_{\pm})_{yj} \, (g_{ky} \pm B_{ky})
\pm \frac{2}{(g_{yy})^2} \, (\omega_{\pm})_{yk} \, (g_{jy} \pm B_{jy})
\Big\} 
\, .
\end{align}
\esubeq
We comment that the tensors $(\hat{N}'_{\pm})^p{}_{mn}$ are neither prepared nor needed, 
because, as defined in (\ref{Nhat}), they are described in terms of contractions of $(J'_{\pm})^m{}_n$ with $\hat{\nabla}'_p (J'_{\pm})^q{}_r$.
We emphasize that the expressions (\ref{hnabla_J_T_}), (\ref{hnabla_omega_T_}), (\ref{N'_}) and (\ref{ScrC'-2}) are complicated, but it seems to have some meaningful structures. 
In Section \ref{sect:integrability} (and Appendix \ref{sect:WS_SUSY}), we will impose appropriate conditions on them, and elaborate geometrical significance under T-duality.

When we consider an almost bi-hypercomplex (\ref{gJomega_a}) and its T-dualized spaces, 
we just attach a subscript $a$ (where $a$ runs $a = 1,2,3$) to $(J_{\pm})^m{}_n$.
The tensors $(\omega_{\pm})_{mn}$, $(N_{\pm})^p{}_{mn}$ and $(\Scr{C}_{\pm})^p{}_{mn}$ also carry this subscript as discussed in \cite{Kimura:2022dma}.

\section{Integrability after T-duality}
\label{sect:integrability}

In this section, we evaluate T-duality from ${\cal M}$ to ${\cal M}'$ in the absence of worldsheet supersymmetry.
In order to extract remarkable features, we introduce some reductions on the most general result in Section \ref{sect:Full}.
Now we consider three cases.
In these cases, we assume the integrability condition of the almost complex structures on ${\cal M}$:
(i) bi-hermitian with isometry in the presence of torsion,
(ii) hermitian with isometry in the absence of torsion, 
(iii) K\"{a}hler with isometry.
Performing T-duality of them,
we investigate whether the almost complex structures are integrable or not on the T-dualized space ${\cal M}'$.

\subsection{Bi-hermitian with isometry in the presence of torsion}
\label{sect:general_H1}

We begin with a bi-hermitian space with isometry. 
We assume that the almost complex structures $(J_{\pm})^m{}_n$ are integrable, i.e.,
the Nijenhuis tensors $(N_{\pm})^p{}_{mn}$ vanish.
On the other hand, we do not impose any conditions on the contorsion tensor $C_{mn}{}^p$. Then the covariant derivatives $\hat{\nabla}_m (J_{\pm})^p{}_n$ and $\Scr{C}$-tensors $(\Scr{C}_{\pm})^p{}_{mn}$ are generically non-zero.
These restrictions are not strong to reduce the property of the T-dualized space ${\cal M}'$:
The form of the T-duality transformations (\ref{hnabla_J_T_}) and (\ref{hnabla_omega_T_}) leads to non-vanishing covariant derivatives on ${\cal M}'$.
Moreover, the Nijenhuis tensors (\ref{N'_}) become nontrivial because the original B-field and the covariant derivatives carry nontrivial values.
The $\Scr{C}$-tensors (\ref{ScrC'-2}) do not describe any useful relations. 
In this system, we just understand that an integrable bi-hermitian space is T-dualized to a non-integrable bi-hermitian space in general.
Caused by the existence of the B-field and non-vanishing covariant derivatives, T-duality breaks integrability of the almost complex structures.
It might be quite interesting if we find a pair of spaces ${\cal M}$ and ${\cal M}'$, where the almost complex structures on ${\cal M}$ are integrable while those on ${\cal M}'$ are not. 
This is because quantum field theory on ${\cal M}$ is controllable by virtue of holomorphy, while that on ${\cal M}'$ is difficult to analyze quantum corrections. 
It is well known that holomorphy is one of the strongest validity both in perturbative and non-perturbative quantum analyses \cite{Seiberg:1994rs, Seiberg:1994bp, Seiberg:1994aj}.
It seems strange that a theory which is well controlled quantum corrections is mapped to another one which is not controlled.
If the above pair exists in physics, there might be an alternative restriction rule on ${\cal M}'$ associated with holomorphy on ${\cal M}$.

In our next consideration, we impose a bit stronger condition on this system.

\subsection{Hermitian with isometry in the absence of torsion}
\label{sect:general_H2}

Next, we consider a hermitian space with isometry in the absence of the B-field.
We again assume that the Nijenhuis tensors vanish and the almost complex structures become integrable.
This is a restricted system from an integrable bi-hermitian space discussed above.
Here we assume that $\Scr{C}$-tensors are generic.

Because we impose that the B-field vanishes, the contorsion tensor in the affine connection also vanishes.
In this case, the geometrical objects are no longer sensitive to worldsheet chirality.
Then the subscripts $\pm$ are omitted, and $(J_{\pm})^m{}_n$, $(\omega_{\pm})_{mn}$, $(N_{\pm})^p{}_{mn}$ and $(\Scr{C}_{\pm})^p{}_{mn}$ are reduced to
$J^m{}_n$, $\omega_{mn}$, $N^p{}_{mn}$ and $\Scr{C}^p{}_{mn}$, respectively.
The obtained geometry is a hermitian space in the absence of torsion.
In this case, the T-dualized objects are described in the following way:
\bsubeq \label{hnabla_J_T_H2}
\begin{align}
\hat{\nabla}'_y (J'_{\pm})^y{}_y
\ &= \ 
0
\, , \\
\hat{\nabla}'_i (J'_{\pm})^y{}_y
\ &= \ 
0
\, , \\
\hat{\nabla}'_y (J'_{\pm})^y{}_k
\ &= \ 
- \frac{1}{g_{yy}} \, g_{ym} \, \nabla_y J^m{}_k
\, , \\
\hat{\nabla}'_i (J'_{\pm})^y{}_k
\ &= \ 
\mp \Big\{
\nabla_i J^m{}_k
- \frac{1}{g_{yy}} \, g_{iy} \, \nabla_y J^m{}_k
\Big\} \, g_{ym}
\, , \\
\hat{\nabla}'_y (J'_{\pm})^j{}_y
\ &= \ 
- \frac{1}{(g_{yy})^2} \, \nabla_y J^j{}_y
\, , \\
\hat{\nabla}'_i (J'_{\pm})^j{}_y
\ &= \ 
\mp \frac{1}{g_{yy}} \, \Big\{
\nabla_i J^j{}_y - \frac{1}{g_{yy}} \, g_{iy} \, \nabla_y J^j{}_y
\Big\}
\, , \\
\hat{\nabla}'_y (J'_{\pm})^j{}_k
\ &= \ 
\pm \frac{1}{g_{yy}} \, \Big\{
\nabla_y J^j{}_k - \frac{1}{g_{yy}} \, g_{ky} \, \nabla_y J^j{}_y 
\Big\}
\, , \\
\hat{\nabla}'_i (J'_{\pm})^j{}_k
\ &= \ 
\Big\{
\nabla_i J^j{}_k - \frac{1}{g_{yy}} \, g_{iy} \, \nabla_y J^j{}_k
\Big\}
- \frac{1}{g_{yy}} \, \Big\{
\nabla_i J^j{}_y - \frac{1}{g_{yy}} \, g_{iy} \, \nabla_y J^j{}_y 
\Big\} \, g_{ky}
\, . 
\end{align}
\esubeq
This implies that the almost complex structure $J^m{}_n$ 
on the original space ${\cal M}$ is split into $(J'_+)^m{}_n$ and $(J'_-)^m{}_n$ on the T-dualized space ${\cal M}'$. 
The covariant derivatives $\hat{\nabla}'_p (J'_{\pm})^m{}_n$ are non-zero if
$\nabla_p J^m{}_n$ are non-zero.
Via the metricity condition (\ref{metricity_3}),
we observe that the covariant derivatives $\hat{\nabla}'_p (\omega'_{\pm})_{mn}$
are also non-zero if $\nabla_p \omega_{mn}$ are non-zero:
\bsubeq \label{hnabla_omega_T_H2}
\begin{align}
\hat{\nabla}'_y (\omega'_{\pm})_{jy}
\ &= \ 
- \frac{1}{(g_{yy})^2} \, \nabla_y \omega_{jy}
\, , \\
\hat{\nabla}'_i (\omega'_{\pm})_{jy}
\ &= \ 
\mp \frac{1}{g_{yy}} \, \nabla_i \omega_{jy}
\pm \frac{1}{(g_{yy})^2} \, g_{iy} \, \nabla_y \omega_{jy}
\, , \\
\hat{\nabla}'_y (\omega'_{\pm})_{jk}
\ &= \ 
\pm \frac{1}{g_{yy}} \, \nabla_y \omega_{jk}
\pm \frac{1}{(g_{yy})^2} \, g_{ky} \, \nabla_y \omega_{yj}
\mp \frac{1}{(g_{yy})^2} \, g_{jy} \, \nabla_y \omega_{yk}
\, , \\
\hat{\nabla}'_i (\omega'_{\pm})_{jk}
\ &= \ 
\nabla_i \omega_{jk}
- \frac{1}{g_{yy}} \, g_{iy} \, \nabla_y \omega_{jk} 
- \frac{1}{g_{yy}} \, g_{jy} \, \nabla_i \omega_{yk} 
- \frac{1}{g_{yy}} \, g_{ky} \, \nabla_i \omega_{jy} 
\nn \\
\ & \ \ \ \ 
+ \frac{1}{(g_{yy})^2} \, g_{iy} \, g_{ky} \, \nabla_y \omega_{jy} 
- \frac{1}{(g_{yy})^2} \, g_{iy} \, g_{jy} \, \nabla_y \omega_{ky} 
\, .
\end{align}
\esubeq
Further, we analyze the Nijenhuis tensors on ${\cal M}'$ in the following way:
\bsubeq \label{N'_H2}
\begin{align}
(N'_{\pm})^y{}_{jy}
\ &= \ 
\frac{3}{g_{yy}} \, \nabla_{[y} \omega_{jm]} \, J^m{}_y
\, , \\
(N'_{\pm})^i{}_{jy}
\ &= \ 
\mp \frac{2}{(g_{yy})^2} \, \nabla_y \omega_{yj} \, J^i{}_y
\, , \\
(N'_{\pm})^y{}_{jk}
\ &= \ 
\mp \nabla_y \omega_{jm} \, J^m{}_k 
\pm \nabla_y \omega_{km} \, J^m{}_j 
\pm \frac{3}{g_{yy}} \, \Big\{
\nabla_{[y} \omega_{jm]} \, g_{yk} 
- \nabla_{[y} \omega_{km]} \, g_{yj}
\Big\} \, J^m{}_y 
\, , \\
(N'_{\pm})^i{}_{jk}
\ &= \ 
- \frac{2}{g_{yy}} \, \Big\{
\nabla_y \omega_{jk} 
+ \frac{1}{g_{yy}} \, g_{jy} \, \nabla_y \omega_{ky} 
- \frac{1}{g_{yy}} \, g_{ky} \, \nabla_y \omega_{jy} 
\Big\} \, J^i{}_y 
\, .
\end{align}
\esubeq
The form of the T-dualized $\Scr{C}$-tensors is still complicated to extract some useful property:
\bsubeq \label{ScrC'-2_H2}
\begin{align}
(\Scr{C}'_{\pm})^y{}_{jy}
\ &= \ 
0
\, , \\
(\Scr{C}'_{\pm})^i{}_{jy}
\ &= \ 
\mp \frac{1}{g_{yy}} \, \Scr{C}^i{}_{jy} 
+ \frac{2}{g_{yy}} \, \nabla_y J^m{}_j \, J^i{}_m 
\nn \\
\ & \ \ \ \
- \frac{2}{(g_{yy})^2} \, \Big\{
\nabla_y J^m{}_y \, J^i{}_m \, g_{jy} 
- \nabla_y J^i{}_y \, \omega_{jy} 
+ \nabla_y \omega_{jy} \, J^i{}_y 
\Big\} 
\, , \\
(\Scr{C}'_{\pm})^y{}_{jk}
\ &= \ 
\mp g_{my} \, \Scr{C}^m{}_{jk} 
- \nabla_y \omega_{mj} \, J^m{}_k 
+ \nabla_y \omega_{mk} \, J^m{}_j 
\nn \\
\ & \ \ \ \ 
- \frac{2}{g_{yy}} \, \nabla_y J^m{}_y \, \Big\{ \omega_{mj} \, g_{ky} - \omega_{mk} \, g_{jy} \Big\} 
+ \frac{2}{g_{yy}} \, \Big\{
\nabla_y \omega_{yk} \, \omega_{yj} 
- \nabla_y \omega_{yj} \, \omega_{yk}
\Big\} 
\, , \\
(\Scr{C}'_{\pm})^i{}_{jk}
\ &= \ 
\Scr{C}^i{}_{jk}
+ \frac{1}{g_{yy}} \, g_{ky} \, \Scr{C}^i{}_{yj} 
- \frac{1}{g_{yy}} \, g_{jy} \, \Scr{C}^i{}_{yk} 
\mp \frac{2}{g_{yy}} \, \Big\{
\nabla_y J^i{}_j \, \omega_{yk}
- \nabla_y J^i{}_k \, \omega_{yj}
- \nabla_y \omega_{jk} \, J^i{}_y
\Big\} 
\nn \\
\ & \ \ \ \ 
\pm \frac{2}{(g_{yy})^2} \, J^i{}_y \, \Big\{ 
\nabla_y \omega_{yj} \, g_{ky} 
- \nabla_y \omega_{yk} \, g_{jy} 
\Big\} 
\mp \frac{2}{(g_{yy})^2} \, \nabla_y J^i{}_y \, 
\Big\{ \omega_{yj} \, g_{ky} - \omega_{yk} \, g_{jy} \Big\} 
\, .
\end{align}
\esubeq
This insists that the T-duality transformation again breaks integrability of the almost complex structures
if $\nabla_m \omega_{np}$ on the original space ${\cal M}$ is non-zero.
We also understand that, if the B-field on the original space ${\cal M}$ is absent, the covariant derivatives are mapped to the sum of the covariant derivatives through T-duality.
We summarize the property of the T-duality transformation of a general hermitian space with isometry in the absence of torsion in Table \ref{table:HvsabH}.
\begin{table}[h]
\tbcaption{T-duality transformation from a general hermitian space with isometry in the absence of torsion.}
\label{table:HvsabH}
\vspace*{-3mm}
\begin{center}
\slb{.85}{\renewcommand{\arraystretch}{1.35}
\begin{tabular}{Wl{80mm}|Wl{80mm}} \hline
$(g,J,\omega)$ on hermitian with isometry ${\cal M}$
& $(g', J'_{\pm}, \omega'_{\pm})$ on T-dualized space ${\cal M}'$
\\ \hline\hline
$g_{mn} = \omega_{mp} \, J^p{}_n$
& 
$g'_{mn} = (\omega'_{\pm})_{mp} \, (J'_{\pm})^p{}_n$
\\
$J^m{}_n = - g^{mp} \, \omega_{pn}$
&
$(J'_{\pm})^m{}_n = - g'{}^{mp} \, (\omega'_{\pm})_{pn}$
\\
$\omega_{mn} = - g_{mp} \, J^p{}_n$
&
$(\omega'_{\pm})_{mn} = - g'_{mp} \, (J'_{\pm})^p{}_n$
\\ \hdashline
$\nabla_m g_{np} = 0$ 
&
$\hat{\nabla}'_m g'_{np} = 0$ 
\\
$\d \omega \neq 0 \Leftrightarrow \nabla_m J^p{}_n \neq 0$
&
$\hat{\nabla}'_m (J'_{\pm})^p{}_n \neq 0 \Leftrightarrow \hat{\nabla}'_m (\omega'_{\pm})_{np} \neq 0$
\\ \hdashline
$N^p{}_{mn} = 0$ 
&
$(N'_{\pm})^p{}_{mn} \neq 0$
\\ \hdashline
$\Scr{C}^p{}_{mn}$ is generic
& 
$(\Scr{C}'_{\pm})^p{}_{mn} \neq 0$
\\ \hline
\end{tabular}
}
\end{center}
\end{table}

Following from the above results, 
we understand
that the non-vanishing covariant derivatives prevent integrability under T-duality.
Then we further restrict that the covariant derivatives of the complex structure on the original space ${\cal M}$ are trivial.
This is nothing but a K\"{a}hler space.

\subsection{K\"{a}hler with isometry}
\label{sect:general_K}

We discuss general property of the T-duality transformation of a K\"{a}hler space with isometry ${\cal M}$.
In particular, we investigate the transformation of the Nijenhuis tensor and $\Scr{C}$-tensor.
We find that integrability of the almost complex structures is preserved even after the T-duality transformation, and the resulting space ${\cal M}'$ is also a complex geometry in the presence of torsion.
The K\"{a}hler condition on the original space ${\cal M}$
is so strong that the (non-)existence of worldsheet supersymmetry does not influence on its geometrical feature seriously.


We introduce the K\"{a}hler condition represented by
\bsubeq \label{Kahler-condition}
\begin{align}
(N_{\pm})^p{}_{mn} \ = \ 0
\, , \quad
\d \omega_{\pm} \ = \ 0
\, . 
\end{align}
Combining this with the metricity condition (\ref{metricity_2}),
we further obtain
\begin{align}
C_{mn}{}^p \ = \ 0
\, , \quad
\hat{\nabla}_m (J_{\pm})^p{}_n
\ = \ 0
\, . \label{hnabla_J_}
\end{align}
We immediately find that the $\Scr{C}$-tensors (\ref{ScrC_def}) vanish:
\begin{align}
(\Scr{C}_{\pm})^p{}_{mn} \ = \ 0
\, .
\end{align}
\esubeq 
Since the contorsion tensor vanishes, the affine connection $\hat{\Gamma}_{mn}{}^p$ (\ref{affine_conn}) is reduced to the Christoffel symbol $\Gamma_{mn}{}^p$ (\ref{Christoffel_conn}). In this case the subscripts ``$\pm$'' in (\ref{Kahler-condition}) are again omitted.


Now we evaluate the T-duality transformation from ${\cal M}$ to ${\cal M}'$.
The Buscher rule (\ref{Buscher-rule}) generates the B-field
and the nontrivial contorsion tensor $C'_{mn}{}^p$ on ${\cal M}'$.
Since the almost complex structure on the original space ${\cal M}$ is covariantly constant (\ref{hnabla_J_}), the T-dualized covariant derivatives (\ref{hnabla_J_T_}) automatically vanish:
\begin{align}
\hat{\nabla}'_m (J'_{\pm})^p{}_n
\ &= \ 
0
\, . \label{hnabla_J_T_K2}
\end{align}
Analogously, the covariant derivatives $\hat{\nabla}'_m (\omega'_{\pm})_{np}$ (\ref{hnabla_omega_T_}) become trivial: 
\begin{align}
\hat{\nabla}'_m (\omega'_{\pm})_{np}
\ &= \ 
0
\, . \label{hnabla_omega_T_K2}
\end{align}
The Nijenhuis tensors $(N'_{\pm})^p{}_{mn}$ (\ref{N'_}) and the $\Scr{C}$-tensors $(\Scr{C}'_{\pm})^p{}_{mn}$ (\ref{ScrC'-2}) also become trivial independently,
even though the contorsion tensor $C'_{mn}{}^p$ on ${\cal M}'$ exists:
\begin{align}
(N'_{\pm})^p{}_{mn} \ = \ 0
\, , \quad
(\Scr{C}'_{\pm})^p{}_{mn} \ = \ 0
\, . \label{N'_T_K2}
\end{align}
This is because the B-field on the original K\"{a}hler space ${\cal M}$ is trivial. 
The vanishing Nijenhuis tensors indicate that the almost complex structures on ${\cal M}'$ are integrable.
Analogously, the vanishing $\Scr{C}$-tensors show that the H-flux $H_{mnp}$ has a following relation
\begin{align}
H'_{mnp}
\ &= \ 
3 \, (J'_{\pm})^q{}_{[m} \, (J'_{\pm})^r{}_n \, H'_{p]qr}
\, . \label{H'_property}
\end{align}
Indeed this relation appears in string compactification scenarios as discussed in \cite{Strominger:1986uh}.


We emphasize that the existence of worldsheet supersymmetry is not important
in this system, because the K\"{a}hler condition is sufficiently strong to set various tensors trivial. 
We summarize the property of the T-duality transformation of a general K\"{a}hler space with isometry in Table \ref{table:KvsbH}.
\begin{table}[h]
\tbcaption{T-duality transformation from a general K\"{a}hler space with isometry.}
\label{table:KvsbH}
\vspace*{-3mm}
\begin{center}
\slb{.85}{\renewcommand{\arraystretch}{1.35}
\begin{tabular}{Wl{80mm}|Wl{80mm}} \hline
$(g,J,\omega)$ on K\"{a}hler with isometry ${\cal M}$
& $(g', J'_{\pm}, \omega'_{\pm})$ on T-dualized space ${\cal M}'$
\\ \hline\hline
$g_{mn} = \omega_{mp} \, J^p{}_n$
& 
$g'_{mn} = (\omega'_{\pm})_{mp} \, (J'_{\pm})^p{}_n$
\\
$J^m{}_n = - g^{mp} \, \omega_{pn}$
&
$(J'_{\pm})^m{}_n = - g'{}^{mp} \, (\omega'_{\pm})_{pn}$
\\
$\omega_{mn} = - g_{mp} \, J^p{}_n$
&
$(\omega'_{\pm})_{mn} = - g'_{mp} \, (J'_{\pm})^p{}_n$
\\ \hdashline
$\nabla_m g_{np} = 0$ 
&
$\hat{\nabla}'_m g'_{np} = 0$ 
\\
$\d \omega = 0 \Leftrightarrow \nabla_m J^p{}_n = 0$ 
&
$\hat{\nabla}'_m (J'_{\pm})^p{}_n = 0 \Leftrightarrow \hat{\nabla}'_m (\omega'_{\pm})_{np} = 0$
\\ \hdashline
$N^p{}_{mn} = 0$ 
&
$(N'_{\pm})^p{}_{mn} = 0$
\\ \hdashline
$\Scr{C}^p{}_{mn} = 0$ 
& 
$(\Scr{C}'_{\pm})^p{}_{mn} = 0$
\\ \hline
\end{tabular}
}
\end{center}
\end{table}

In Section \ref{sect:examples}, we demonstrate T-duality 
of two typical examples: $\mathbb{C}P^2$ and Taub-NUT.
In both cases, the integrability condition is preserved under T-duality,
and the $\Scr{C}$-tensors impose 
the appropriate property (\ref{H'_property}) on the contorsion tensor.

\subsection{Examples of K\"{a}hler with isometry}
\label{sect:examples}

In this subsection, we demonstrate T-duality of two typical K\"{a}hler spaces with isometry.
In the first example, we consider T-duality of the complex projective space $\mathbb{C}P^2$. This manifold appears as a useful toy model in string theory.
Second, we concretely describe the Taub-NUT space, which gives rise to a Kaluza-Klein (KK) monopole in string theory. 
We perform T-duality transformations and obtain stringy geometries such as an H-monopole and an exotic $5^2_2$-brane background,  as discussed in the literature.

\subsubsection{Complex projective space $\mathbb{C}P^2$ and its T-duality}
\label{sect:CP2}

Complex projective spaces are compact K\"{a}hler manifolds with isometry.
Although it is difficult to apply themselves to the string sigma model with conformal symmetry, because they are not Ricci-flat,
it is interesting to explore their T-dualized spaces as parts of full geometries \cite{Bakhmatov:2010fp, Bouwknegt:2003wp}.
Here we focus on the two-dimensional complex projective space $\mathbb{C}P^2$\footnote{It is impossible to define a spinor on the $\mathbb{C}P^2$ space. Although we do not touch this issue in the present work, it is interesting to discuss carefully whether spinors could (not) be defined on the T-dualized space.}.

\subsubsection*{Geometrical aspect of $\mathbb{C}P^2$}

The geometrical aspect of $\mathbb{C}P^2$ has been well studied in the literature.
We first write down the K\"{a}hler potential
\begin{align}
K (\zeta^{\sf a},\ol{\zeta}{}^{\ol{\sf b}}) 
\ &= \ 
\frac{1}{c} \, \log \Big[ 1 + c \, \big( |\zeta^1|^2 + |\zeta^2|^2 \big) \Big]
\, , 
\end{align}
where $c$ is a positive constant and $(\zeta^{\sf a}, \ol{\zeta}{}^{\ol{\sf b}})$ are the complex local (inhomogeneous) coordinates (${\sf a} = 1,2$).
The K\"{a}hler potential $K$ 
gives the Fubini-Study metric $g_{{\sf a} \ol{\sf b}}$ in the following way:
\begin{align}
\d s^2 
\ &= \
g_{{\sf a} \ol{\sf b}} \, \d \zeta^{\sf a} \, \d \ol{\zeta}{}^{\ol{\sf b}}
\ = \ 
\frac{\del^2 K}{\del \zeta^{\sf a} \del \ol{\zeta}{}^{\ol{\sf b}}} \, \d \zeta^{\sf a} \, \d \ol{\zeta}{}^{\ol{\sf b}}
\nn \\
\ &= \ 
\frac{|\d \zeta^1|^2 + |\d \zeta^2|^2
+ c \, |\d (\zeta^1 \ol{\zeta}{}^{\ol{2}})|^2 
- c \, \d (|\zeta^1|^2) \, \d (|\zeta^2|^2)}{(1 + c (|\zeta^1|^2 + |\zeta^2|^2))^2}
\, . \label{CP2_ds2_cpx}
\end{align}
We often express the K\"{a}hler metric as 
$g_{{\sf a} \ol{\sf b}} = \del_{\sf a} \del_{\ol{\sf b}} K \eqqcolon K_{{\sf a} \ol{\sf b}}$.

In order to express the isometry direction explicitly,
we transform the complex coordinates 
$(\zeta^{\sf a}, \ol{\zeta}{}^{\ol{\sf b}})$ (${\sf a},\ol{\sf b} = 1,2$) 
to the real polar coordinates
$\x^m = (r, \t, \varphi, \psi)^{\T}$ 
in such a way that
\begin{alignat}{2}
\zeta^1 \ &= \ 
r \cos \tfrac{\t}{2} \, \e^{\frac{\I}{2} (\psi + \varphi)}
\, , &\qquad
\zeta^2 \ &= \ 
r \sin \tfrac{\t}{2} \, \e^{\frac{\I}{2} (\psi - \varphi)}
\, , \label{CP2_polar}
\end{alignat}
where the domain of the radial coordinate is 
$0 \leq r < \infty$, 
while those of the three angular coordinates are
$0 \leq \t \leq \pi$, 
$0 \leq \varphi \leq 2 \pi$, and
$0 \leq \psi \leq 4 \pi$, respectively \cite{Gibbons:1978zy}\footnote{The polar coordinates are related to the 
${\rm SU(2)}$ left-invariant one-forms $\sigma_i$ ($i = x,y,z$) \cite{Eguchi:1978gw}.}.
The coordinate transformation from the complex coordinates $\zeta^{\sf a}$ to the real coordinates $\x^m$ yields the transformation of the metric in such a way that
\bsubeq \label{CP2_ds2_polar}
\begin{gather}
\d s^2 
\ = \ 
g_{{\sf a} \ol{\sf b}} \, \d \zeta^{\sf a} \, \d \ol{\zeta}{}^{\ol{\sf b}}
\ = \ 
g_{{\sf a} \ol{\sf b}} \, \frac{\del \zeta^{\sf a}}{\del \x^{(m}} \, \frac{\del \ol{\zeta}{}^{\ol{\sf b}}}{\del \x^{n)}} 
\, \d \x^m \, \d \x^n
\ = \ 
g_{mn} \, \d \x^m \, \d \x^n
\, , \\
g_{mn} 
\ = \ 
\half \Big\{
\frac{\del \zeta^{\sf a}}{\del \x^m} \, \frac{\del \ol{\zeta}{}^{\ol{\sf b}}}{\del \x^n} 
+ \frac{\del \zeta^{\sf a}}{\del \x^n} \, \frac{\del \ol{\zeta}{}^{\ol{\sf b}}}{\del \x^m}
\Big\} \, g_{{\sf a} \ol{\sf b}}
\, . 
\end{gather}
\esubeq
By using this, the Fubini-Study metric (\ref{CP2_ds2_cpx}) in the finite region of $r$ is given as 
\begin{align}
\d s^2 
\ &= \ 
\frac{\d r^2}{(1 + c r^2)^2} 
+ \frac{r^2}{4 (1 + c r^2)^2} \big( \d \psi + \cos \t \, \d \varphi \big)^2
+ \frac{r^2}{4 (1 + c r^2)} \big( \d \t^2 + \sin^2 \t \, \d \varphi^2 \big)
\, . \label{ds2_polar2}
\end{align}
When $r$ goes to infinity, 
we further introduce a new coordinate $u\coloneqq 1/r$ and rewrite (\ref{ds2_polar2}) as
\begin{align}
\d s^2 
\ &= \ 
\frac{\d u^2}{(u^2 + c)^2} 
+ \frac{u^2}{4 (u^2 + c)^2} \big( \d \psi + \cos \t \, \d \varphi \big)^2
+ \frac{1}{4 (u^2 + c)} \big( \d \t^2 + \sin^2 \t \, \d \varphi^2 \big)
\, . \label{CP2_ds2_polar_inf}
\end{align}
For later convenience,
we introduce the function $f(r) \coloneqq 1 + c r^2 = \e^{cK}$
and express the metric (\ref{ds2_polar2}) in the matrix form:
\begin{align}
g_{mn} \ &= \ 
\left(
{\renewcommand{\arraystretch}{.9}
\begin{array}{cccc}
g_{rr} 
& g_{r\t} 
& g_{r \varphi} 
& g_{r \psi} 
\\
g_{\t r} 
& g_{\t \t} 
& g_{\t \varphi} 
& g_{\t \psi} 
\\
g_{\varphi r} 
& g_{\varphi \t} 
& g_{\varphi \varphi}
& g_{\varphi \psi} 
\\
g_{\psi r} 
& g_{\psi \t} 
& g_{\psi \varphi} 
& g_{\psi \psi} 
\end{array}
}
\right)
\ = \ 
\left(
{\renewcommand{\arraystretch}{.9}
\begin{array}{cccc}
\frac{1}{f^2} 
& 0 
& 0 
& 0 
\\
0 
& \frac{r^2}{4 f} 
& 0 
& 0 
\\
0 
& 0 
& \frac{r^2 (1 + cr^2 \sin^2 \t)}{4 f^2} 
& \frac{r^2 \cos \t}{4 f^2} 
\\
0 
& 0 
& \frac{r^2 \cos \t}{4 f^2} 
& \frac{r^2}{4 f^2} 
\end{array}
}
\right)
\, . \label{CP2g_polar_matrix}
\end{align}
It is clear that both (\ref{ds2_polar2}) and (\ref{CP2_ds2_polar_inf})
do not depend on the angular coordinate $\psi$.
Indeed, the location of $\psi = 4 \pi$ coincides with that of $\psi = 0$,
i.e., the $\psi$-direction is periodic under the translation $\psi \to \psi + 4 \pi$.
This indicates that there exists isometry along the $\psi$-direction.

Here we describe the almost complex structure on $\mathbb{C}P^2$.
In each local patch, we can express its components as
$J^{\sf a}{}_{\sf b} = \I \, \delta^{\sf a}_{\sf b}$ 
(and its complex conjugate
$J^{\ol{\sf a}}{}_{\ol{\sf b}} = - \I \, \delta^{\ol{\sf a}}_{\ol{\sf b}}$).
Under the coordinate transformation from the complex coordinates to the real coordinates, 
the almost complex structure is transformed in the following way:
\begin{align}
J^m{}_n \ &= \ 
\frac{\del \x^m}{\del \zeta^{\sf a}} \frac{\del \zeta^{\sf b}}{\del \x^n} J^{\sf a}{}_{\sf b} 
+ \frac{\del \x^m}{\del \ol{\zeta}{}^{\ol{\sf a}}} \frac{\del \ol{\zeta}{}^{\ol{\sf b}}}{\del \x^n} J^{\ol{\sf a}}{}_{\ol{\sf b}}
\, . \label{def_J}
\end{align}
Applying (\ref{CP2_polar}) to this transformation,
we obtain the real components of the almost complex structure whose matrix form is
\begin{align}
J^m{}_n
\ &= \ 
\left(
{\renewcommand{\arraystretch}{.9}
\begin{array}{cccc}
J^r{}_r 
& J^r{}_{\t} 
& J^r{}_{\varphi} 
& J^r{}_{\psi} 
\\
J^{\t}{}_r 
& J^{\t}{}_{\t} 
& J^{\t}{}_{\varphi} 
& J^{\t}{}_{\psi} 
\\
J^{\varphi}{}_r 
& J^{\varphi}{}_{\t} 
& J^{\varphi}{}_{\varphi} 
& J^{\varphi}{}_{\psi} 
\\
J^{\psi}{}_r 
& J^{\psi}{}_{\t} 
& J^{\psi}{}_{\varphi} 
& J^{\psi}{}_{\psi} 
\end{array}
}
\right)
\ = \
\left(
{\renewcommand{\arraystretch}{.9}
\begin{array}{cccc}
0 
& 0 
& - \frac{r}{2} \cos \t 
& - \frac{r}{2} 
\\
0 
& 0 
& \sin \t 
& 0 
\\
0 
& - \csc \t 
& 0 
& 0 
\\
\frac{2}{r} 
& \cot \t 
& 0 
& 0 
\end{array}
}
\right)
\, . \label{CP2J_polar_matrix}
\end{align}
Of course this expression keeps the property $J^m{}_p J^p{}_n = - \delta^m_n$.

In the case of the fundamental two-form,
the transformation from the complex coordinates to the real coordinates is given by
\bsubeq \label{def_omega}
\begin{gather}
\omega 
\ = \ 
\half \, \omega_{{\sf a} \ol{\sf b}} \, \d \zeta^{\sf a} \w \d \ol{\zeta}{}^{\ol{\sf b}}
\ = \ 
\half \, \omega_{{\sf a} \ol{\sf b}} \, \frac{\del \zeta^{\sf a}}{\del \x^{[m}} \, \frac{\del \ol{\zeta}{}^{\ol{\sf b}}}{\del \x^{n]}} 
\, \d \x^m \w \d \x^n
\ = \ 
\frac{1}{2} \, \omega_{mn} \, \d \x^m \w \d \x^n
\, , \\
\omega_{mn} 
\ = \ 
\half \Big\{
\frac{\del \zeta^{\sf a}}{\del \x^m} \, \frac{\del \ol{\zeta}{}^{\ol{\sf b}}}{\del \x^n} 
- \frac{\del \zeta^{\sf a}}{\del \x^n} \, \frac{\del \ol{\zeta}{}^{\ol{\sf b}}}{\del \x^m}
\Big\} \, (\I K_{{\sf a} \ol{\sf b}})
\, , 
\end{gather}
\esubeq
where we used the fact that the fundamental two-form can be represented by the K\"{a}hler form $\omega_{{\sf a} \ol{\sf b}} = \I K_{\sf {a} \ol{\sf b}}$.
We obtain the components of the fundamental two-form whose matrix form is 
\begin{align}
\omega_{mn}
\ &= \ 
\left(
{\renewcommand{\arraystretch}{.9}
\begin{array}{cccc}
\omega_{rr} 
& \omega_{r \t} 
& \omega_{r \varphi} 
& \omega_{r \psi} 
\\
\omega_{\t r} 
& \omega_{\t \t} 
& \omega_{\t \varphi} 
& \omega_{\t \psi} 
\\
\omega_{\varphi r} 
& \omega_{\varphi \t} 
& \omega_{\varphi \varphi}
& \omega_{\varphi \psi} 
\\
\omega_{\psi r} 
& \omega_{\psi \t} 
& \omega_{\psi \varphi} 
& \omega_{\psi \psi} 
\end{array}
}
\right)
\ = \
\left(
{\renewcommand{\arraystretch}{.9}
\begin{array}{cccc}
0
& 0 
& \frac{r \cos \t}{2f^2}
& \frac{r}{2f^2}
\\
0 
& 0 
& - \frac{r^2 \sin \t}{4 f}
& 0
\\
- \frac{r \cos \t}{2f^2} 
& \frac{r^2 \sin \t}{4f}
& 0
& 0
\\
- \frac{r}{2 f^2}
& 0
& 0
& 0
\end{array}
}
\right)
\, . \label{CP2omega_polar_matrix}
\end{align}

Since $\mathbb{C}P^2$ is K\"{a}hler, 
the set of 
the metric (\ref{CP2g_polar_matrix}), the almost complex structure (\ref{CP2J_polar_matrix}),
and the fundamental two-form (\ref{CP2omega_polar_matrix}) 
satisfies the almost hermitian structure (\ref{AHS_property}).
Since the fundamental two-form is closed, there are no torsion terms.
Furthermore, since the almost complex structure is covariantly constant with respect to the Christoffel symbol,
the Nijenhuis tensor vanishes.

\subsubsection*{T-duality along $\psi$-direction}

We perform the T-duality transformation along the $\psi$-direction 
of the complex projective space $\mathbb{C}P^2$.
We are interested in the contorsion tensor, the affine connection, and the integrability condition on the T-dualized space.
The Buscher rule (\ref{Buscher-rule}) mixes the metric, B-field and dilaton.
Analogously, the rule (\ref{KSS-rule}) mixes the almost complex structure and fundamental two-form.
Here, for simplicity, we choose the B-field and dilaton on $\mathbb{C}P^2$ to be trivial.
Then the T-duality transformation of the almost hermitian structure is described.

First, we exhibit the metric after the T-duality transformation by the Buscher rule (\ref{Buscher-rule}):
\bsubeq \label{g_T-CP2}
\begin{align}
g'_{mn}
\ &= \ 
\slb{.9}{$
\left(
\begin{array}{cccc}
g'_{rr}
& g'_{r \t}
& g'_{r \varphi}
& g'_{r \tpsi}
\\
g'_{\t r}
& g'_{\t \t}
& g'_{\t \varphi}
& g'_{\t \tpsi}
\\
g'_{\varphi r}
& g'_{\varphi \t}
& g'_{\varphi \varphi}
& g'_{\varphi \tpsi}
\\
g'_{\tpsi r}
& g'_{\tpsi \t}
& g'_{\tpsi \varphi}
& g'_{\tpsi \tpsi}
\end{array}
\right)
$}
\ = \ 
\slb{.9}{$ 
\left(
\begin{array}{cccc}
\frac{1}{f^2}
& 0 
& 0 
& 0 
\\
0 
& \frac{r^2}{4f}
& 0 
& 0 
\\
0 
& 0 
& \frac{r^2 \sin^2 \t}{4f}
& 0 
\\
0 
& 0 
& 0 
& \frac{4f^2}{r^2}
\end{array}
\right)
$}
\, .
\end{align}
\esubeq
The B-field on the T-dualized space is generated by the Buscher rule: 
\begin{align}
B'_{mn}
\ &= \ 
\slb{.9}{$
\left(
\begin{array}{cccc}
B'_{rr}
& B'_{r \t}
& B'_{r \varphi}
& B'_{r \tpsi}
\\
B'_{\t r}
& B'_{\t \t}
& B'_{\t \varphi}
& B'_{\t \tpsi}
\\
B'_{\varphi r}
& B'_{\varphi \t}
& B'_{\varphi \varphi}
& B'_{\varphi \tpsi}
\\
B'_{\tpsi r}
& B'_{\tpsi \t}
& B'_{\tpsi \varphi}
& B'_{\tpsi \tpsi}
\end{array}
\right)
$}
\ = \ 
\slb{.9}{$ 
\left(
\begin{array}{cccc}
 0 & 0 & 0 & 0 \\
 0 & 0 & 0 & 0 \\
 0 & 0 & 0 & \cos \t \\
 0 & 0 & - \cos \t & 0
\end{array}
\right)
$}
\, . \label{B_T-CP2}
\end{align}
The dilaton is generated as $\phi' = \half \log (4 f^2/r^2)$, which we do not investigate in this work.
By using the rule (\ref{KSS-rule}), we obtain the explicit form of the T-dualized almost complex structures
\begin{align}
(J'_{\pm})^m{}_n
\ &= \ 
\slb{.9}{$
\left(
\begin{array}{cccc}
(J'_{\pm})^{r}{}_{r}
& (J'_{\pm})^{r}{}_{\t}
& (J'_{\pm})^{r}{}_{\varphi}
& (J'_{\pm})^{r}{}_{\tpsi}
\\
(J'_{\pm})^{\t}{}_{r}
& (J'_{\pm})^{\t}{}_{\t}
& (J'_{\pm})^{\t}{}_{\varphi}
& (J'_{\pm})^{\t}{}_{\tpsi}
\\
(J'_{\pm})^{\varphi}{}_{r}
& (J'_{\pm})^{\varphi}{}_{\t}
& (J'_{\pm})^{\varphi}{}_{\varphi}
& (J'_{\pm})^{\varphi}{}_{\tpsi}
\\
(J'_{\pm})^{\tpsi}{}_{r}
& (J'_{\pm})^{\tpsi}{}_{\t}
& (J'_{\pm})^{\tpsi}{}_{\varphi}
& (J'_{\pm})^{\tpsi}{}_{\tpsi}
\end{array}
\right)
$}
\ = \ 
\slb{.9}{$ 
\left(
\begin{array}{cccc}
0 
& 0 
& 0 
& \pm \frac{2f^2}{r} 
\\
0 
& 0 
& \sin \t
& 0 
\\
0 
& - \csc \t
& 0 
& 0 
\\
\mp \frac{r}{2f^2}
& 0 
& 0 
& 0
\end{array}
\right)
$}
\, . \label{J_T-CP2}
\end{align}
We can check that the commuting property of the almost complex structures is preserved. 
In the same way, we obtain the explicit form of the fundamental two-forms
\begin{align}
(\omega'_{\pm})_{mn}
\ &= \
\slb{.9}{$
\left(
\begin{array}{cccc}
(\omega'_{\pm})_{rr}
& (\omega'_{\pm})_{r \t}
& (\omega'_{\pm})_{r \varphi}
& (\omega'_{\pm})_{r \tpsi}
\\
(\omega'_{\pm})_{\t r}
& (\omega'_{\pm})_{\t \t}
& (\omega'_{\pm})_{\t \varphi}
& (\omega'_{\pm})_{\t \tpsi}
\\
(\omega'_{\pm})_{\varphi r}
& (\omega'_{\pm})_{\varphi \t}
& (\omega'_{\pm})_{\varphi \varphi}
& (\omega'_{\pm})_{\varphi \tpsi}
\\
(\omega'_{\pm})_{\tpsi r}
& (\omega'_{\pm})_{\tpsi \t}
& (\omega'_{\pm})_{\tpsi \varphi}
& (\omega'_{\pm})_{\tpsi \tpsi}
\end{array}
\right)
$}
\ = \ 
\slb{.9}{$ 
\left(
\begin{array}{cccc}
0 
& 0 
& 0 
& \mp \frac{2}{r} 
\\
0 
& 0 
& - \frac{r^2 \sin \t}{4f}
& 0 
\\
0 
& \frac{r^2 \sin \t}{4f}
& 0 
& 0 
\\
\pm \frac{2}{r} 
& 0 
& 0 
& 0
\end{array}
\right)
$}
\, . \label{Omega_T-CP2}
\end{align}
Notice that they are no longer closed $\d \omega'_{\pm} \neq 0$.

We would like to explore integrability of the almost complex structures (\ref{J_T-CP2}).
The Christoffel symbol is given in terms of the T-dualized metric (\ref{g_T-CP2}) as follows:
\bsubeq \label{Christoffel_T-CP2}
\begin{alignat}{2}
\Gamma'_{mn}{}^p 
\ &= \ 
\slb{.9}{$ 
\left(
\begin{array}{cccc}
\Gamma'_{rr}{}^p
& \Gamma'_{r \t}{}^p
& \Gamma'_{r \varphi}{}^p
& \Gamma'_{r \tpsi}{}^p
\\
\Gamma'_{\t r}{}^p
& \Gamma'_{\t \t}{}^p
& \Gamma'_{\t \varphi}{}^p
& \Gamma'_{\t \tpsi}{}^p
\\
\Gamma'_{\varphi r}{}^p
& \Gamma'_{\varphi \t}{}^p
& \Gamma'_{\varphi \varphi}{}^p
& \Gamma'_{\varphi \tpsi}{}^p
\\
\Gamma'_{\tpsi r}{}^p
& \Gamma'_{\tpsi \t}{}^p
& \Gamma'_{\tpsi \varphi}{}^p
& \Gamma'_{\tpsi \tpsi}{}^p
\end{array}
\right)
$}
\, , \\
\Gamma'_{mn}{}^r
\ &= \
\slb{.9}{$ 
\left(
\begin{array}{cccc}
- \frac{2cr}{f}
& 0 
& 0 
& 0 
\\
0 
& - \frac{r}{4} 
& 0 
& 0 
\\
0 
& 0 
& - \frac{r \sin^2 \t}{4}
& 0 
\\
0 
& 0 
& 0 
& \frac{4f^3 (1 - cr^2)}{r^3}
\end{array}
\right)
$}
\, , &\quad
\Gamma'_{mn}{}^{\t}
\ &= \
\slb{.9}{$ 
\left(
\begin{array}{cccc}
0 
& \frac{1}{rf}
& 0 
& 0 
\\
\frac{1}{rf}
& 0 
& 0 
& 0 
\\
0 
& 0 
& - \cos \t \sin \t
& 0 
\\
0 
& 0 
& 0 
& 0
\end{array}
\right)
$}
\, , \\
\Gamma'_{mn}{}^{\varphi}
\ &= \
\slb{.9}{$ 
\left(
\begin{array}{cccc}
0 
& 0 
& \frac{1}{rf}
& 0 
\\
0 
& 0 
& \cot \t
& 0 
\\
\frac{1}{rf}
& \cot \t
& 0 
& 0 
\\
0 
& 0 
& 0 
& 0 
\end{array}
\right)
$}
\, , &\quad 
\Gamma'_{mn}{}^{\tpsi}
\ &= \
\slb{.9}{$ 
\left(
\begin{array}{cccc}
0 
& 0 
& 0 
& \frac{-1+cr^2}{rf}
\\
0 
& 0 
& 0 
& 0 
\\
0 
& 0 
& 0 
& 0 
\\
\frac{-1+cr^2}{rf}
& 0 
& 0 
& 0
\end{array}
\right)
$}
\, .
\end{alignat}
\esubeq
The field-strength $H'_{mnp} = 3 \del'_{[m} B'_{np]}$ yields a tensor $C'_{mn}{}^p$ by contracting with the inverse metric $g'{}^{mn}$ as expressed in (\ref{H-flux}):
\bsubeq \label{contorsion_T-CP2}
\begin{alignat}{2}
C'_{mn}{}^p
\ &= \ 
\slb{.9}{$ 
\left(
\begin{array}{cccc}
C'_{rr}{}^p
& C'_{r \t}{}^p
& C'_{r \varphi}{}^p
& C'_{r \tpsi}{}^p
\\
C'_{\t r}{}^p
& C'_{\t \t}{}^p
& C'_{\t \varphi}{}^p
& C'_{\t \tpsi}{}^p
\\
C'_{\varphi r}{}^p
& C'_{\varphi \t}{}^p
& C'_{\varphi \varphi}{}^p
& C'_{\varphi \tpsi}{}^p
\\
C'_{\tpsi r}{}^p
& C'_{\tpsi \t}{}^p
& C'_{\tpsi \varphi}{}^p
& C'_{\tpsi \tpsi}{}^p
\end{array}
\right)
$}
\, , \\
C'_{mn}{}^r
\ &= \ 
0
\, , &\quad
C'_{mn}{}^{\t}
\ &= \ 
\slb{.9}{$ 
\left(
\begin{array}{cccc}
0 
& 0 
& 0 
& 0 
\\
0 
& 0 
& 0 
& 0 
\\
0 
& 0 
& 0 
& -\frac{4f \sin \t}{r^2} 
\\
0 
& 0 
& \frac{4f \sin \t}{r^2} 
& 0
\end{array}
\right)
$}
\, , \\
C'_{mn}{}^{\varphi}
\ &= \ 
\slb{.9}{$ 
\left(
\begin{array}{cccc}
0 
& 0 
& 0 
& 0 
\\
0 
& 0 
& 0 
& \frac{4f \csc \t}{r^2}
\\
0 
& 0 
& 0 
& 0 
\\
0 
& -\frac{4f \csc \t}{r^2}
& 0 
& 0
\end{array}
\right)
$}
\, , &\quad
C'_{mn}{}^{\tpsi}
\ &= \ 
\slb{.9}{$ 
\left(
\begin{array}{cccc}
0 
& 0 
& 0 
& 0 \\
0 
& 0 
& -\frac{r^2 \sin \t}{4f^2}
& 0 
\\
0 
& \frac{r^2 \sin \t}{4f^2}
& 0 
& 0 
\\
0 
& 0 
& 0 
& 0
\end{array}
\right)
$}
\, .
\end{alignat}
\esubeq
This is regarded as the contorsion tensor on the T-dualized space.
Combining this with the Christoffel symbol (\ref{Christoffel_T-CP2}),
we describe the affine connection $\hat{\Gamma}'_{mn}{}^p$ (see, for details, Appendix \ref{sect:T-dual_basic}). 
With respect to this affine connection, we evaluate the covariant derivative of the almost complex structures (\ref{J_T-CP2}):
\begin{align}
\hat{\nabla}'_m (J'_{\pm})^p{}_n
\ &= \ 
0
\, . \label{nablahat_J} 
\end{align}
Analogously, we evaluate the covariant derivative of the fundamental two-forms (\ref{Omega_T-CP2}):
\begin{align}
\hat{\nabla}'_m (\omega'_{\pm})_{np}
\ &= \ 
0
\, . \label{nablahat_omega}
\end{align}
We also observe that the Nijenhuis tensors and $\Scr{C}$-tensors vanish:
\begin{align}
(N'_{\pm})^p{}_{mn}
\ &= \ 
0
\, , \quad 
(\Scr{C}'_{\pm})^p{}_{mn}
\ = \ 
0 
\, . \label{Nijenhuis_ScrC_T-CP2} 
\end{align}
This implies that the almost complex structures (\ref{J_T-CP2}) are integrable even after the T-duality transformation.
Hence we conclude that the T-dualized space along the $\psi$-direction of $\mathbb{C}P^2$ is a complex space with contorsion (\ref{contorsion_T-CP2}), where the almost complex structures are covariantly constant.
This is a typical example of non-K\"{a}hler geometries with isometry.

\subsubsection{Taub-NUT and its T-duality}
\label{sect:TN}

In this part, we consider a hyper-K\"{a}hler space with isometry and its T-dualized systems.
A hyper-K\"{a}hler space has a set of three integrable complex structures $J_a$ without torsion, where the subscript runs $a = 1,2,3$.
Then the associated fundamental two-forms $\omega_a$ are all closed $\d \omega_a = 0$.
In the literature, 
string sigma models with worldsheet supersymmetry on KK-monopoles and H-monopoles \cite{Tong:2002rq, Harvey:2005ab, Okuyama:2005gx} have been deeply investigated.
In an analogous way, an exotic $5^2_2$-brane from KK-monopoles has been also analyzed (see, for instance,  \cite{deBoer:2012ma} and references therein). 
A gauged linear sigma model on this exotic brane \cite{Kimura:2013fda} is 
successfully constructed in terms of two-dimensional complex chiral superfields, 
where integrability of the background space of the exotic brane is implicitly admitted.
Once the complex superfield formalism is applicable,
we can analyze its worldsheet instanton corrections \cite{Kimura:2013zva, Kimura:2018hph, Kimura:2022jyp} following from \cite{Tong:2002rq, Harvey:2005ab, Okuyama:2005gx}. 
This is a typical example which represents the existence of the integrability condition implicitly.

\subsubsection*{Taub-NUT space (KK-monopole)}

We introduce the Taub-NUT space given by four-dimensional space with coordinates $x^m$ ($m = 1,2,3,4$).
This space is a noncompact hyper-K\"{a}hler, which is characterized by the metric $g_{mn}$, complex structures $J_a$ and fundamental two-forms $\omega_a$ ($a = 1,2,3$).
On this space the B-field $B_{mn}$ and dilaton $\phi$ are trivial.
They are expressed in the following way:
\begin{gather}
\d s^2 
\ = \ 
H \, \d x_{123}^2 
+ H^{-1} \big( \d x^4 + A_i \, \d x^i \big)^2
\, , \quad
B_{mn} \ = \ 0
\, , \ls 
\e^{2\phi} \ = \ 1
\, . \label{Taub-NUT}
\end{gather}
We introduce a harmonic function $H$ and a vector potential $A_i$ on the flat three-dimensional subspace expanded by $x^i$ ($i = 1,2,3$). 
They are given by 
\begin{gather}
H \ = \ 
1 + \frac{\ell_0}{\sqrt{2} \, r}
\, , \quad
r^2 \ = \ \sum_{i=1}^3 (x^i)^2
\, , \quad
\del_{[i} A_{j]} \ = \ \frac{1}{2!} \, \eps_{ijk} \, \del_k H
\, , \label{monopole_eq}
\end{gather}
where $\ell_0$ is a constant.
We notice that $\eps_{ijk}$ is the Levi-Civita symbol on the three-dimensional subspace.
Since the B-field is trivial, its field-strength is also trivial $C_{mn}{}^p = 0$.
The matrix form of the metric $g_{mn}$ is useful in later discussions:
\begin{align}
g_{mn} \ = \
\left(
\begin{array}{cccc}
g_{11} & g_{12} & g_{13} & g_{14} \\
g_{21} & g_{22} & g_{23} & g_{24} \\
g_{31} & g_{32} & g_{33} & g_{34} \\
g_{41} & g_{42} & g_{43} & g_{44} 
\end{array}
\right)
\ = \ 
\slb{.85}{$
\dfrac{1}{H} 
\left(
\begin{array}{cccc}
H^2 + (A_1)^2 & A_1 A_2 & A_1 A_3 & A_1 \\
A_1 A_2 & H^2 + (A_2)^2 & A_2 A_3 & A_2 \\
A_1 A_3 & A_2 A_3 & H^2 + (A_3)^2 & A_3 \\
A_1 & A_2 & A_3 & 1
\end{array}
\right)
$}
\, . \label{TN_g_matrix}
\end{align}
Analogously, we also express the matrix form of the complex structures 
\bsubeq \label{CS_TN}
\begin{align}
(J_a)^m{}_n
\ &= \ 
\left(
\begin{array}{cccc}
(J_a)^1{}_1 & (J_a)^1{}_2 & (J_a)^1{}_3 & (J_a)^1{}_4 \\
(J_a)^2{}_1 & (J_a)^2{}_2 & (J_a)^2{}_3 & (J_a)^2{}_4 \\
(J_a)^3{}_1 & (J_a)^3{}_2 & (J_a)^3{}_3 & (J_a)^3{}_4 \\
(J_a)^4{}_1 & (J_a)^4{}_2 & (J_a)^4{}_3 & (J_a)^4{}_4 
\end{array}
\right)
\, , 
\end{align}
whose expressions are given as
\begin{align}
(J_1)^m{}_n \ &= \
\slb{.85}{$
- \dfrac{1}{H}
\left(
\begin{array}{cccc}
A_1 & A_2 & A_3 & 1 \\
0 & 0 & H & 0 \\
0 & - H & 0 & 0 \\
- H^2 - (A_1)^2 & -A_1 A_2 + A_3 H & - A_1 A_3 - A_2 H & - A_1 
\end{array}
\right)
$}
\, , \\
(J_2)^m{}_n \ &= \ 
\slb{.85}{$
- \dfrac{1}{H}
\left(
\begin{array}{cccc}
0 & 0 & -H & 0 \\
A_1 & A_2 & A_3 & 1 \\
H & 0 & 0 & 0 \\
-A_1 A_2 - A_3 H & - H^2 - (A_2)^2 & -A_2 A_3 + A_1 H & - A_2
\end{array}
\right)
$}
\, , \\
(J_3)^m{}_n \ &= \ 
\slb{.85}{$
- \dfrac{1}{H}
\left(
\begin{array}{cccc}
0 & H & 0 & 0 \\
- H & 0 & 0 & 0 \\
A_1 & A_2 & A_3 & 1 \\
-A_1 A_3 + A_2 H & -A_2 A_3 - A_1 H & - H^2 - (A_3)^2 & - A_3
\end{array}
\right)
$}
\, , 
\end{align}
\esubeq
and the matrix form of the fundamental two-forms $\omega_a$:
\bsubeq \label{FS_TN}
\begin{align}
(\omega_1)_{mn}
\ &= \
\slb{.85}{$
\left(
\begin{array}{cccc}
(\omega_1)_{11} & (\omega_1)_{12} & (\omega_1)_{13} & (\omega_1)_{14} \\
(\omega_1)_{21} & (\omega_1)_{22} & (\omega_1)_{23} & (\omega_1)_{24} \\
(\omega_1)_{31} & (\omega_1)_{32} & (\omega_1)_{33} & (\omega_1)_{34} \\
(\omega_1)_{41} & (\omega_1)_{42} & (\omega_1)_{43} & (\omega_1)_{44} 
\end{array}
\right)
$} 
\ = \ 
\slb{.85}{$
\left(
\begin{array}{cccc}
0 & A_2 & A_3 & 1 \\
-A_2 & 0 & H & 0 \\
-A_3 & -H & 0 & 0 \\
-1 & 0 & 0 & 0
\end{array}
\right)
$}
\, , \\
(\omega_2)_{mn} \ &= \
\slb{.85}{$
\left(
\begin{array}{cccc}
0 & -A_1 & -H & 0 \\
A_1 & 0 & A_3 & 1 \\
H & -A_3 & 0 & 0 \\
0 & -1 & 0 & 0 
\end{array}
\right)
$}
\, , \\
(\omega_3)_{mn} \ &= \
\slb{.85}{$
\left(
\begin{array}{cccc}
0 & H & -A_1 & 0 \\
-H & 0 & -A_2 & 0 \\
A_1 & A_2 & 0 & 1 \\
0 & 0 & -1 & 0
\end{array}
\right)
$}
\, .
\end{align}
\esubeq
Due to (\ref{monopole_eq}), we immediately find that the fundamental two-forms are closed $\d \omega_a = 0$.
We also find that the Nijenhuis tensors $(N_a)^p{}_{mn}$, associated with the respective complex structures $J_a$ via (\ref{N}), vanish.
Hence we understand that all $J_a$ are integrable.

\subsubsection*{H-monopole (smeared NS5-brane)}

Let us perform the T-duality transformation along the isometry direction $y=4$, 
in order to describe the H-monopole (or, called the smeared NS5-brane) background data ($g'$, $J'_{a,\pm}$, $\omega'_{a,\pm}$).
The metric, B-field and dilaton are described as
\begin{gather}
g'_{mn} \ = \
\slb{.85}{$
\left(
\begin{array}{cccc}
H & 0 & 0 & 0 \\
0 & H & 0 & 0 \\
0 & 0 & H & 0 \\
0 & 0 & 0 & H
\end{array}
\right)
$}
\, , \quad
B'_{mn} \ = \ 
\slb{.85}{$
\left(
\begin{array}{cccc}
0 & 0 & 0 & A_1 \\
0 & 0 & 0 & A_2 \\
0 & 0 & 0 & A_3 \\
-A_1 & -A_2 & -A_3 & 0 
\end{array}
\right)
$}
\, , \quad 
\e^{2 \phi'} \ = \ H
\, . \label{H-monopole}
\end{gather}
Each complex stricture $J_a$ on the Taub-NUT space is split into 
$J'_{a,\pm}$, 
which satisfy the commuting condition $[J'_{a,+}, J'_{a,-}] = 0$,
in the following way:
\bsubeq \label{CS_Hmono}
\begin{alignat}{2}
(J'_{1,+})^m{}_n \ &= \
\slb{.85}{$
\left(
\begin{array}{cccc}
0 & 0 & 0 & 1 \\
0 & 0 & -1 & 0 \\
0 & 1 & 0 & 0 \\
-1 & 0 & 0 & 0 
\end{array}
\right)
$}
\, , &\quad 
(J'_{1,-})^m{}_n \ &= \
\slb{.85}{$
\left(
\begin{array}{cccc}
0 & 0 & 0 & -1 \\
0 & 0 & -1 & 0 \\
0 & 1 & 0 & 0 \\
1 & 0 & 0 & 0 
\end{array}
\right)
$}
\, , \\
(J'_{2,+})^m{}_n \ &= \ 
\slb{.85}{$
\left(
\begin{array}{cccc}
0 & 0 & 1 & 0 \\
0 & 0 & 0 & 1 \\
-1 & 0 & 0 & 0 \\
0 & -1 & 0 & 0 
\end{array}
\right)
$}
\, , &\quad 
(J'_{2,-})^m{}_n \ &= \
\slb{.85}{$
\left(
\begin{array}{cccc}
0 & 0 & 1 & 0 \\
0 & 0 & 0 & -1 \\
-1 & 0 & 0 & 0 \\
0 & 1 & 0 & 0 
\end{array}
\right)
$}
\, , \\
(J'_{3,+})^m{}_n \ &= \ 
\slb{.85}{$
\left(
\begin{array}{cccc}
0 & -1 & 0 & 0 \\
1 & 0 & 0 & 0 \\
0 & 0 & 0 & 1 \\
0 & 0 & -1 & 0 
\end{array}
\right)
$}
\, , &\quad 
(J'_{3,-})^m{}_n \ &= \
\slb{.85}{$
\left(
\begin{array}{cccc}
0 & -1 & 0 & 0 \\
1 & 0 & 0 & 0 \\
0 & 0 & 0 & -1 \\
0 & 0 & 1 & 0 
\end{array}
\right)
$}
\, .
\end{alignat}
\esubeq
The fundamental two-forms are also split as follows:
\bsubeq \label{FS_Hmono}
\begin{alignat}{2}
(\omega'_{1,+})_{mn} \ &= \ 
\slb{.85}{$
\left(
\begin{array}{cccc}
0 & 0 & 0 & -H \\
0 & 0 & H & 0 \\
0 & -H & 0 & 0 \\
H & 0 & 0 & 0 
\end{array}
\right)
$}
\, , &\quad 
(\omega'_{1,-})_{mn} \ &= \
\slb{.85}{$
\left(
\begin{array}{cccc}
0 & 0 & 0 & H \\
0 & 0 & H & 0 \\
0 & -H & 0 & 0 \\
-H & 0 & 0 & 0 
\end{array}
\right)
$}
\, , \\
(\omega'_{2,+})_{mn} \ &= \
\slb{.85}{$
\left(
\begin{array}{cccc}
0 & 0 & -H & 0 \\
0 & 0 & 0 & -H \\
H & 0 & 0 & 0 \\
0 & H & 0 & 0 
\end{array}
\right)
$}
\, , &\quad 
(\omega'_{2,-})_{mn} \ &= \
\slb{.85}{$
\left(
\begin{array}{cccc}
0 & 0 & -H & 0 \\
0 & 0 & 0 & H \\
H & 0 & 0 & 0 \\
0 & -H & 0 & 0 
\end{array}
\right)
$}
\, , \\
(\omega'_{3,+})_{mn} \ &= \
\slb{.85}{$
\left(
\begin{array}{cccc}
0 & H & 0 & 0 \\
-H & 0 & 0 & 0 \\
0 & 0 & 0 & -H \\
0 & 0 & H & 0 
\end{array}
\right)
$}
\, , &\quad 
(\omega'_{3,-})_{mn} \ &= \
\slb{.85}{$
\left(
\begin{array}{cccc}
0 & H & 0 & 0 \\
-H & 0 & 0 & 0 \\
0 & 0 & 0 & H \\
0 & 0 & -H & 0 
\end{array}
\right)
$}
\, .
\end{alignat}
\esubeq
Since the almost complex structures (\ref{CS_Hmono}) are covariantly constant (\ref{hnabla_J_T_K2}) in the presence of torsion, worldsheet supersymmetry can be introduced \cite{Gates:1984nk}. Furthermore, the vanishing Nijenhuis tensors (\ref{N'_T_K2}) indicate that complex-valued superfields can be utilized to construct the string sigma model for the H-monopole \cite{Tong:2002rq}.

\subsubsection*{Exotic $5^2_2$-brane}

Here we go back to the original Taub-NUT space (\ref{Taub-NUT}).
Smearing the $3$-direction of the Taub-NUT space, and performing the T-duality transformation along this direction, 
we obtain the exotic $5^2_2$-brane background data ($\tilde{g}$, $\tilde{J}_{a,\pm}$, $\tilde{\omega}_{a,\pm}$) with two isometry directions $m = 3,4$.
The metric, B-field and dilaton are nontrivially described as
\bsubeq \label{522_cartesian}
\begin{gather}
\tilde{g}_{mn} \ = \
\slb{.85}{$
\left(
\begin{array}{cccc}
H & 0 & 0 & 0 \\
0 & H & 0 & 0 \\
0 & 0 & H/K & 0 \\
0 & 0 & 0 & H/K
\end{array}
\right)
$}
\, , \quad
\tilde{B}_{mn} \ = \ 
\slb{.85}{$
\left(
\begin{array}{cccc}
0 & 0 & 0 & 0 \\
0 & 0 & 0 & 0 \\
0 & 0 & 0 & - A_3/K \\
0 & 0 & A_3/K & 0 
\end{array}
\right)
$}
\, , \quad 
\e^{2\tilde{\phi}} \ = \ \frac{H}{K}
\, , \\
H \ = \ 
h_0 + \sigma \log \Big( \frac{\mu}{\r} \Big)
\, , \quad
A_3 \ = - \sigma \t
\, , \quad
K \ = \ H^2 + (A_3)^2
\, , \\
\r^2 \ = \ (x^1)^2 + (x^2)^2
\, , \quad
\t \ = \ \arctan \Big( \frac{x^2}{x^1} \Big)
\, ,
\end{gather}
\esubeq
where $h_0$, $\sigma$ and $\mu$ are constants, while $\r$ and $\t$ are the radial and angular coordinates on the $(x^1,x^2)$-plane.
Analogous to the H-monopole system, the complex structures $J_a$ are split into $\tilde{J}_{a,\pm}$ in the following way:
\bsubeq \label{CS_522}
\begin{alignat}{2}
(\tilde{J}_{1,+})^m{}_n \ &= \
\slb{.85}{$
\left(
\begin{array}{cccc}
0 & 0 & A_3/K & -H/K \\
0 & 0 & H/K & A_3/K \\
-A_3 & -H & 0 & 0 \\
H & -A_3 & 0 & 0
\end{array}
\right)
$}
\, , &\quad 
(\tilde{J}_{1,-})^m{}_n \ &= \
\slb{.85}{$
\left(
\begin{array}{cccc}
0 & 0 & -A_3/K & -H/K \\
0 & 0 & -H/K & A_3/K \\
A_3 & H & 0 & 0 \\
H & -A_3 & 0 & 0
\end{array}
\right)
$}
\, , \\
(\tilde{J}_{2,+})^m{}_n \ &= \ 
\slb{.85}{$
\left(
\begin{array}{cccc}
0 & 0 & -H/K & -A_3/K \\
0 & 0 & A_3/K & -H/K \\
H & -A_3 & 0 & 0 \\
A_3 & H & 0 & 0
\end{array}
\right)
$}
\, , &\quad 
(\tilde{J}_{2,-})^m{}_n \ &= \
\slb{.85}{$
\left(
\begin{array}{cccc}
0 & 0 & H/K & -A_3/K \\
0 & 0 & -A_3/K & -H/K \\
-H & A_3 & 0 & 0 \\
A_3 & H & 0 & 0
\end{array}
\right)
$}
\, , \\
(\tilde{J}_{3,+})^m{}_n \ &= \ 
\slb{.85}{$
\left(
\begin{array}{cccc}
0 & -1 & 0 & 0 \\
1 & 0 & 0 & 0 \\
0 & 0 & 0 & 1 \\
0 & 0 & -1 & 0
\end{array}
\right)
$}
\, , &\quad 
(\tilde{J}_{3,-})^m{}_n \ &= \
\slb{.85}{$
\left(
\begin{array}{cccc}
0 & -1 & 0 & 0 \\
1 & 0 & 0 & 0 \\
0 & 0 & 0 & -1 \\
0 & 0 & 1 & 0
\end{array}
\right)
$}
\, .
\end{alignat}
\esubeq
These almost complex structures commute with each other $[\tilde{J}_{a,+}, \tilde{J}_{a,-}] = 0$ and are integrable \cite{Kimura:2022dma}. 
This commuting property is consistent with that of the gauged linear sigma model for the exotic five-brane description \cite{Kimura:2013fda}, where the model is written in terms of $\N=(2,2)$ chiral and twisted chiral superfields (or their relaxed types \cite{Kimura:2018ain}), rather than $\N=(4,4)$ superfields \cite{Hassan:1994mq, Bakas:1995hc}\footnote{In this article, we do not explore a description in the worldsheet $\N=(4,4)$ superfield formalism.}.
Associated with these complex structures, the fundamental two-forms are also given by
\bsubeq \label{FS_522}
\begin{alignat}{2}
(\tilde{\omega}_{1,+})_{mn} \ &= \ 
\slb{.85}{$
\dfrac{H}{K}
\left(
\begin{array}{cccc}
0 & 0 & -A_3 & H \\
0 & 0 & -H & -A_3 \\
A_3 & H & 0 & 0 \\
-H & A_3 & 0 & 0 
\end{array}
\right)
$}
\, , &\quad 
(\tilde{\omega}_{1,-})_{mn} \ &= \
\slb{.85}{$
\dfrac{H}{K}
\left(
\begin{array}{cccc}
0 & 0 & A_3 & H \\
0 & 0 & H & -A_3 \\
-A_3 & -H & 0 & 0 \\
-H & A_3 & 0 & 0 
\end{array}
\right)
$}
\, , \\
(\tilde{\omega}_{2,+})_{mn} \ &= \
\slb{.85}{$
\dfrac{H}{K}
\left(
\begin{array}{cccc}
0 & 0 & H & A_3 \\
0 & 0 & -A_3 & H \\
-H & A_3 & 0 & 0 \\
-A_3 & -H & 0 & 0 
\end{array}
\right)
$}
\, , &\quad 
(\tilde{\omega}_{2,-})_{mn} \ &= \
\slb{.85}{$
\dfrac{H}{K}
\left(
\begin{array}{cccc}
0 & 0 & -H & A_3 \\
0 & 0 & A_3 & H \\
H & -A_3 & 0 & 0 \\
-A_3 & -H & 0 & 0 
\end{array}
\right)
$}
\, , \\
(\tilde{\omega}_{3,+})_{mn} \ &= \
\slb{.85}{$
\dfrac{H}{K}
\left(
\begin{array}{cccc}
0 & K & 0 & 0 \\
-K & 0 & 0 & 0 \\
0 & 0 & 0 & -1 \\
0 & 0 & 1 & 0
\end{array}
\right)
$}
\, , &\quad 
(\tilde{\omega}_{3,-})_{mn} \ &= \
\slb{.85}{$
\dfrac{H}{K}
\left(
\begin{array}{cccc}
0 & K & 0 & 0 \\
-K & 0 & 0 & 0 \\
0 & 0 & 0 & 1 \\
0 & 0 & -1 & 0
\end{array}
\right)
$}
\, .
\end{alignat}
\esubeq
This result is newly appeared 
because we did not discuss the smearing procedure to generate a new isometry in our previous work \cite{Kimura:2022dma}.
Here we show 
that our T-duality manipulation is also applicable to a geometry with multiple isometry directions.
By concrete analysis, we immediately find that the almost complex structures (\ref{CS_522}) are still integrable, even after the smearing procedure. 
Notice that they are subject to a nontrivial monodromy structure occurred by the angular coordinate shift $\t \to \t + 2 \pi$ on the $(x^1,x^2)$-plane 
because $K$ is shifted $H^2 \to H^2 + (2 \pi \sigma)^2$.

\section{Conclusion and discussions}
\label{sect:conclusion}

In this article, we analyzed T-duality transformation on almost bi-hermitian spaces with isometry, in which two almost complex structures commute with each other.
First, we wrote down all of the components of geometrical objects after T-duality transformation. 
Next, we discussed target spaces of string sigma models and introduced certain constraints on them.
In the main discussion, 
we investigated the T-duality transformation of various tensors in the absence of worldsheet supersymmetry.
In a general hermitian space, T-duality breaks integrability of the almost complex structures.
As a special case, however, if we imposed the K\"{a}hler condition on ${\cal M}$,
its T-dualized space ${\cal M}'$ still admits integrability of the almost complex structures.
We summarized our results in Table \ref{table:KvsbH} (see, as comparisons, Table \ref{table:HvsabH} in Section \ref{sect:general_H2} and Table \ref{table:bHvsbH} in Appendix \ref{sect:WS_SUSY}).
As typical examples, 
we demonstrated the T-duality behavior of the complex projective space $\mathbb{C}P^2$ and Taub-NUT space, 
where the existence of worldsheet supersymmetry is not a serious issue.

By showing the explicit transformation rule for the Nijenhuis tensor, 
we found the relation between the integrability and the covariant constancy conditions 
on the almost complex structures $J^m{}_n$ in the T-dualized frame. 
Although these conditions are automatically satisfied in supersymmetric cases, 
we stress that it is not evident whether they continue to hold in the absence of worldsheet supersymmetry.
In most literature, the integrability and the covariant constancy of $J^m{}_n$ is implicitly guaranteed by string worldsheet supersymmetry.

We concentrated on behaviors of almost complex structures.
As an application, 
we would like to investigate the almost complex structures on the background geometry of the various defect $(p,q)$ five-branes \cite{Kimura:2014wga, Kimura:2014bea}.
There also exist other geometrical objects \cite{Ivanov:2000ai} such as 
(i) the Bismut torsion, which describes the property of torsion;
(ii) the Riemann curvature with torsion, which is directly connected to the Einstein equation;
(iii) the Lee-form, which relates the torsion and dilaton in the geometrical viewpoint, 
and so forth.
It would be interesting to investigate them in the framework of conventional geometry and of DFT, as discussed in \cite{Jeon:2011cn, Hohm:2011si, Kimura:2023nvt}.

In this article, Abelian T-duality associated with a translational Killing symmetry is considered.
There exists another type of T-duality associated with a rotational Killing symmetry, which appears in $\N=4$ extended worldsheet supersymmetry \cite{Bakas:1995hc, Hassan:1995je, Alvarez:1995ai}. 
When this type of T-duality is considered, the description in Section \ref{sect:Full} should be modified.

It would be also important to study a space with non-commuting almost complex structures \cite{Ivanov:1994ec, Lindstrom:2004hi, Goteman:2009ye} (and 
\cite{Ivanov:1995jb, Ivanov:1995yp}\footnote{In these two papers the harmonic superspace technique is utilized.})
when torsionful geometry is classified and understood completely.
Such a geometry is suitable with a sigma model given by semi-chiral superfields \cite{Buscher:1987uw, Ivanov:1994ec, Lindstrom:2004hi, Lindstrom:2022nto, Hull:2024gvy}. 
They are also applied to a geometry with monodromy \cite{Kimura:2015qze, Kimura:2018ain}, for instance.
If the considering space possesses vanishing Nijenhuis tensor(s), 
there exist integrable complex structure(s), 
and it becomes guaranteed that the string sigma model can be described in terms of {\it complex-valued} superfields such as chiral, twisted chiral, and/or semi-chiral superfields.
We will study the relation between integrability of non-commuting almost complex structures and T-duality in future works.

\subsubsection*{Acknowledgments}

The authors would like to thank the Yukawa Institute for Theoretical Physics at Kyoto University.
Discussions during the YITP workshop {\tt YITP-W-24-08} on ``Strings and Fields 2024'' were useful to complete this work. 
This work is supported in part by Grant-in-Aid for Scientific Research (C), JSPS KAKENHI Grant Numbers {\tt JP23K03398} (TK).

\begin{appendix}
\subsection*{Appendix}

\section{T-duality transformation of affine connection}
\label{sect:T-dual_basic}

In this appendix, we generally write down the T-duality transformation rule of the affine connection on an almost bi-hermitian space. 
All of them can be derived in terms of the Buscher rule (\ref{Buscher-rule}) and (\ref{KSS-rule}).
We refer to the T-dualized Christoffel symbol, contorsion tensor, and affine connection as $\Gamma'_{mn}{}^p$, $C'_{mn}{}^p$ and $\hat{\Gamma}'_{mn}{}^p$, respectively.

First, we express the Christoffel symbol. All components are given as follows:
\bsubeq \label{Gamma'}
\begin{align}
\Gamma'_{yy}{}^y
\ &= \ 
\frac{1}{(g_{yy})^2} \, \Gamma_{yy}{}^m \, B_{my} 
\, , \\
\Gamma'_{yy}{}^k
\ &= \ 
- \frac{1}{(g_{yy})^2} \, \Gamma_{yy}{}^k 
\, , \\
\Gamma'_{iy}{}^y
\ &= \ 
\frac{1}{2 g_{yy}} C_{iy}{}^m \, B_{my} 
+ \frac{1}{(g_{yy})^2} \Gamma_{yy}{}^m \, ( B_{my} \, B_{iy} + g_{yy} \, g_{im} )
\, , \\
\Gamma'_{iy}{}^k
\ &= \ 
- \frac{1}{2 g_{yy}} C_{iy}{}^k
- \frac{1}{(g_{yy})^2} \, \Gamma_{yy}{}^k \, B_{iy} 
\, , \\
\Gamma'_{ij}{}^y
\ &= \ 
- \Gamma_{ij}{}^m \, B_{my}
+ \frac{1}{g_{yy}} \, \Gamma_{iy}{}^m \, B_{my} \, g_{jy}
+ \frac{1}{g_{yy}} \, \Gamma_{jy}{}^m \, B_{my} \, g_{iy}
+ \frac{1}{g_{yy}} \, \Gamma_{yy}{}^m \, ( g_{mi} \, B_{jy} + g_{mj} \, B_{iy} )
\nn \\
\ & \ \ \ \ 
- \frac{1}{(g_{yy})^2} \, \Gamma_{yy}{}^m \, B_{my} \, ( g_{iy} \, g_{jy} - B_{iy} \, B_{jy} )
\nn \\
\ & \ \ \ \ 
+ \frac{1}{2} \, ( \del_i B_{jy} + \del_j B_{iy} )
- \frac{1}{2 g_{yy}} \, B_{my} \, ( C_{yi}{}^m \, B_{jy} + C_{yj}{}^m \, B_{iy} )
\, , \\
\Gamma'_{ij}{}^k
\ &= \ 
\Gamma_{ij}{}^k 
- \frac{1}{g_{yy}} \, \Gamma_{iy}{}^k \, g_{jy} 
- \frac{1}{g_{yy}} \, \Gamma_{jy}{}^k \, g_{iy}
+ \frac{1}{(g_{yy})^2} \, \Gamma_{yy}{}^k \, ( g_{iy} \, g_{jy} - B_{iy} \, B_{jy} )
\nn \\
\ & \ \ \ \ 
+ \frac{1}{2 g_{yy}} \, ( C_{yi}{}^k \, B_{jy} + C_{yj}{}^k \, B_{iy} )
\, .
\end{align}
\esubeq
Next, we exhibit the transformation rule of the contorsion tensor.
The components are described in the following way: 
\bsubeq \label{C'_mnp}
\begin{align}
C'_{iy}{}^y
\ &= \ 
\frac{2}{(g_{yy})^2} \, ( \Gamma_{iy}{}^m \, g_{yy} - \Gamma_{yy}{}^m \, g_{iy} ) \, B_{my} 
\, , \\
C'_{iy}{}^k
\ &= \ 
- \frac{2}{(g_{yy})^2} \, ( \Gamma_{iy}{}^k \, g_{yy} - \Gamma_{yy}{}^k \, g_{iy} )
\, , \\
C'_{ij}{}^y
\ &= \ 
- C_{ij}{}^m \, B_{my} 
+ \frac{1}{g_{yy}} \, C_{yj}{}^m \, B_{my} \, g_{iy} 
- \frac{1}{g_{yy}} \, C_{yi}{}^m \, B_{my} \, g_{jy} 
\nn \\
\ & \ \ \ \ 
- \Gamma_{jy}{}^m \, \big( g_{mi} + \frac{2}{g_{yy}} \, B_{my} \, B_{iy} \big)
+ \Gamma_{iy}{}^m \, \big( g_{mj} + \frac{2}{g_{yy}} \, B_{my} \, B_{jy} \big)
\nn \\
\ & \ \ \ \
+ \frac{2}{g_{yy}} \, \Gamma_{yy}{}^m \, g_{jy} \, \big( g_{mi} + \frac{1}{g_{yy}} \, B_{my} \, B_{iy} \big)
- \frac{2}{g_{yy}} \, \Gamma_{yy}{}^m \, g_{iy} \, \big( g_{mj} + \frac{1}{g_{yy}} \, B_{my} \, B_{jy} \big)
\, , \\
C'_{ij}{}^k
\ &= \ 
C_{ij}{}^k
- \frac{1}{g_{yy}} ( g_{iy} \, C_{yj}{}^k + g_{jy} \, C_{iy}{}^k ) 
\nn \\
\ & \ \ \ \ 
- \frac{2}{(g_{yy})^2} \, \Big\{
( \Gamma_{iy}{}^k \, B_{jy} - \Gamma_{jy}{}^k \, B_{iy} ) \, g_{yy} 
+ \Gamma_{yy}{}^k \, ( B_{iy} \, g_{jy} - g_{iy} \, B_{jy} )
\Big\}
\, .
\end{align}
\esubeq
Combining the above two descriptions,
we obtain the T-duality transformation rule of the affine connection with torsion.
The explicit form of the components is written as follows:
\bsubeq \label{hGamma'}
\begin{align}
\hat{\Gamma}'_{yy}{}^y 
\ &= \ 
\frac{1}{(g_{yy})^2} \, \Gamma_{yy}{}^m \, B_{my} 
\, , \\
\hat{\Gamma}'_{yy}{}^k 
\ &= \ 
- \frac{1}{(g_{yy})^2} \, \Gamma_{yy}{}^k 
\, , \\
\hat{\Gamma}'_{iy}{}^y 
\ &= \ 
\frac{1}{g_{yy}} \, \Gamma_{yy}{}^m \, g_{mi} 
+ \frac{2 \xi}{g_{yy}} \, \big( \Gamma_{iy}{}^m + \frac{1}{4 \xi} \, C_{iy}{}^m \big) \, B_{my}
- \frac{2 \xi}{(g_{yy})^2} \, \Gamma_{yy}{}^m \, \big( g_{iy} - \frac{1}{2 \xi} \, B_{iy} \big) \, B_{my}
\, , \\
\hat{\Gamma}'_{iy}{}^k 
\ &= \ 
- \frac{2 \xi}{g_{yy}} \big( \Gamma_{iy}{}^k + \frac{1}{4 \xi} \, C_{iy}{}^k \big)
+ \frac{2 \xi}{(g_{yy})^2} \, \Gamma_{yy}{}^k \, \big( g_{iy} - \frac{1}{2 \xi} \, B_{iy} \big)
\, , \\
\hat{\Gamma}'_{ij}{}^y 
\ &= \ 
- ( \Gamma_{ij}{}^m + \xi \, C_{ij}{}^m ) \, B_{my} 
+ \xi \, \big( \Gamma_{iy}{}^m + \frac{1}{4 \xi} \, C_{iy}{}^m \big) \, g_{mj} 
- \xi \, \big( \Gamma_{yj}{}^m + \frac{1}{4 \xi} \, C_{yj}{}^m \big) \, g_{mi} 
\nn \\
\ & \ \ \ \ 
+ \frac{1}{g_{yy}} \, ( \Gamma_{iy}{}^m + \xi \, C_{iy}{}^m ) \, B_{my} \, g_{jy} 
+ \frac{1}{g_{yy}} \, ( \Gamma_{yj}{}^m + \xi \, C_{yj}{}^m ) \, B_{my} \, g_{iy} 
\nn \\
\ & \ \ \ \
+ \frac{2 \xi}{g_{yy}} \, \big( \Gamma_{iy}{}^m + \frac{1}{4 \xi} \, C_{iy}{}^m \big) \, B_{my} \, B_{jy} 
- \frac{2 \xi}{g_{yy}} \, \big( \Gamma_{yj}{}^m + \frac{1}{4 \xi} \, C_{yj}{}^m \big) \, B_{my} \, B_{iy} 
\nn \\
\ & \ \ \ \
+ \frac{2 \xi}{g_{yy}} \, \Gamma_{yy}{}^m \, g_{mi} \, \big( g_{jy} + \frac{1}{2 \xi} \, B_{jy} \big)
- \frac{2 \xi}{g_{yy}} \, \Gamma_{yy}{}^m \, g_{mj} \, \big( g_{iy} - \frac{1}{2 \xi} \, B_{iy} \big)
\nn \\
\ & \ \ \ \ 
- \frac{1}{(g_{yy})^2} \, \Gamma_{yy}{}^m \, B_{my} \, g_{iy} \, ( g_{jy} + 2 \xi \, B_{jy} ) 
+ \frac{2 \xi}{(g_{yy})^2} \, \Gamma_{yy}{}^m \, B_{my} \, B_{iy} \, \big( g_{jy} + \frac{1}{2 \xi} \, B_{jy} \big) 
\nn \\
\ & \ \ \ \ 
+ \frac{1}{2} \, ( \del_i B_{jy} + \del_j B_{iy} )
\, , \\
\hat{\Gamma}'_{ij}{}^k
\ &= \ 
( \Gamma_{ij}{}^k + \xi \, C_{ij}{}^k )
- \frac{1}{g_{yy}} \, ( \Gamma_{iy}{}^k + \xi \, C_{iy}{}^k ) \, g_{jy} 
- \frac{1}{g_{yy}} \, ( \Gamma_{yj}{}^k  + \xi \, C_{yj}{}^k ) \, g_{iy}
\nn \\
\ & \ \ \ \ 
- \frac{2 \xi}{g_{yy}} \, \big( \Gamma_{iy}{}^k + \frac{1}{4 \xi} \, C_{iy}{}^k \big) \, B_{jy} 
+ \frac{2 \xi}{g_{yy}} \, \big( \Gamma_{yj}{}^k + \frac{1}{4 \xi} \, C_{yj}{}^k \big) \, B_{iy} 
\nn \\
\ & \ \ \ \ 
+ \frac{1}{(g_{yy})^2} \, \Gamma_{yy}{}^k \, g_{iy} \, ( g_{jy} + 2 \xi \, B_{jy} )
- \frac{2 \xi}{(g_{yy})^2} \, \Gamma_{yy}{}^k \, B_{iy} \, \big( g_{jy} + \frac{1}{2 \xi} \, B_{jy} \big)
\, . 
\end{align}
\esubeq
In the above expression we formally used the constant $\xi$. 
This has been already
set to $\pm \half$ (\ref{constants_fixed}) when the affine connection acts on geometrical objects with subscripts $\pm$ which represent the worldsheet chirality (or holomorphy).

\section{Existing worldsheet supersymmetry}
\label{sect:WS_SUSY}

In this appendix, we briefly review various target spaces of string sigma models with worldsheet supersymmetry as discussed in \cite{Gates:1984nk, deWit:1988fk}\footnote{We focus only on rigid supersymmetry. Note that local supersymmetry \cite{deWit:1988fk} introduces an additional term in the affine connection (\ref{affine_conn}).}. 
In this case, the torsionful covariant derivatives of almost complex structures vanish\footnote{Covariant constancy is sometimes satisfied even in the absence of worldsheet supersymmetry. For instance, a bosonic sigma model whose target space is K\"{a}hler is a typical example, as in Section \ref{sect:general_K}.}.
We consider a general discussion without any geometrical restrictions in the presence of worldsheet supersymmetry.
We prove that the integrability condition of almost complex structures is preserved under T-duality.
We also discuss the Bismut torsion appearing in various string flux compactification scenarios.


Consider a string sigma model with supersymmetry. 
As discussed in \cite{Gates:1984nk, deWit:1988fk},
the worldsheet supersymmetry invariance yields 
\begin{align}
\hat{\nabla}_m (J_{\pm})^n{}_p
\ &= \ 0
\, . \label{hnabla_J_fix_SUSY}
\end{align}
This also implies $\hat{\nabla}_m (\omega_{\pm})_{np} = 0$ by the metricity condition (\ref{metricity_2}).
We thus immediately find that the covariantized $(\hat{N}_{\pm})^p{}_{mn}$ (\ref{Nhat}) vanishes:
\begin{align}
(\hat{N}_{\pm})^p{}_{mn} \ &= \ 0
\, . \label{Nhat_fix_SUSY}
\end{align}
Since the worldsheet supersymmetry is preserved under 
Abelian (or translational)\footref{rotational_T-dual}
T-duality transformation \cite{Hassan:1994mq, Kimura:2022dma},
$\hat{\nabla}'_m (J'_{\pm})^n{}_p$ (\ref{hnabla_J_T_}) 
should vanish. 
Then the tensors $(\hat{N}'_{\pm})^p{}_{mn}$ vanish automatically. 
By using the metricity condition (\ref{metricity_3}), 
$\hat{\nabla}'_m (\omega_{\pm})_{np}$ (\ref{hnabla_omega_T_}) should also vanish.
Furthermore, if there exist two commuting supersymmetries acting on the considering string sigma model Lagrangian $\Scr{L}$, 
i.e., if $\delta$ and $\hat{\delta}$ denote worldsheet supersymmetry variations and satisfy $[\delta, \hat{\delta}] \Scr{L} = 0$ \cite{deWit:1988fk}\footnote{If there exists only one supersymmetry, $\hat{\delta}$ coincides with $\delta$ and the commutator is trivially zero.}, 
the $\Scr{C}$-tensors before T-duality vanish:
\begin{align}
(\Scr{C}_{\pm})^p{}_{mn}
\ &= \ 
0
\, . \label{ScrC_fix_SUSY}
\end{align}
This indicates that the $\Scr{C}$-tensors after the T-duality transformation (\ref{ScrC'-2}) vanish.
Substituting (\ref{Nhat_fix_SUSY}) and (\ref{ScrC_fix_SUSY}) into
(\ref{ScrC_def}), we find that the Nijenhuis tensors $(N_{\pm})^p{}_{mn}$ before the T-duality transformation vanish.
Combining the above results with (\ref{N'_}), we finally obtain the result that the Nijenhuis tensors $(N'_{\pm})^p{}_{mn}$ after the T-duality transformation also vanish:
\begin{align}
(N'_{\pm})^p{}_{mn} \ &= \ 
\mp (\Scr{C}'_{\pm})^p{}_{mn}
\ = \ 0
\, . \label{N'_fix_SUSY}
\end{align}
Therefore, we conclude that the almost complex structures $(J'_{\pm})^m{}_n$ on the T-dualized space ${\cal M}'$ are integrable, and we can introduce complex-valued coordinates on it.
This guarantees that complex superfields in two-dimensional worldsheet sigma model can be mapped to different complex superfields via the Legendre transformation \cite{Rocek:1991ps, Witten:1993yc, Hori:2000kt} without breaking integrability.

We have a comment that,  
if we analyze a supersymmetric string sigma model on a target space with property (\ref{hnabla_J_fix_SUSY}) and (\ref{ScrC_fix_SUSY}),
the Bismut torsion plays a central role in string flux compactification scenarios \cite{Strominger:1986uh, Ivanov:2000ai, Kimura:2006af}.
The definition of the Bismut torsion $T^{\text{(B)}}_{mnp}$ is given as
\begin{align}
(T^{\text{(B)}}_{\pm})_{mnp}
\ \coloneqq& \ \
\frac{3}{2} \, (J_{\pm})^q{}_{[m} \, (J_{\pm})^r{}_{n} \, (J_{\pm})^s{}_{p]} \, \del_q (\omega_{\pm})_{rs}
\nn \\
\ =& \ \
\pm \half \, H_{mnp}
\pm \half \, g_{[m|q|} \, (\Scr{C}_{\pm})^q{}_{np]}
+ \frac{3}{2} \, (J_{\pm})^q{}_{[m} \, (J_{\pm})^r{}_{n} \, (J_{\pm})^s{}_{p]} \, \hat{\nabla}_q (\omega_{\pm})_{rs}
\, . \label{Bismut}
\end{align}
In the presence of worldsheet supersymmetry, we conclude that (\ref{Bismut}) coincides with the H-flux:
\begin{align}
(T^{\text{(B)}}_{\pm})_{mnp}
\ &= \ 
\pm \half \, H_{mnp}
\, , \quad
H_{mnp}
\ = \ 
3 \, (J_{\pm})^q{}_{[m} \, (J_{\pm})^r{}_n \, H_{p]qr}
\, . \label{Bismut_fix_SUSY}
\end{align}
We summarize the above result in Table \ref{table:bHvsbH}.
\begin{table}[h]
\tbcaption{T-duality transformation from an almost bi-hermitian space in the presence of worldsheet supersymmetry.}
\label{table:bHvsbH}
\vspace*{-3mm}
\begin{center}
\slb{.85}{\renewcommand{\arraystretch}{1.35}
\begin{tabular}{Wl{90mm}|Wl{90mm}} \hline
$(g,J_{\pm},\omega_{\pm})$ on almost bi-hermitian ${\cal M}$ 
& $(g', J'_{\pm}, \omega'_{\pm})$ on T-dualized space ${\cal M}'$
\\ \hline\hline
$g_{mn} = (\omega_{\pm})_{mp} \, (J_{\pm})^p{}_n$
& 
$g'_{mn} = (\omega'_{\pm})_{mp} \, (J'_{\pm})^p{}_n$
\\
$(J_{\pm})^m{}_n = - g^{mp} \, (\omega_{\pm})_{pn}$
&
$(J'_{\pm})^m{}_n = - g'{}^{mp} \, (\omega'_{\pm})_{pn}$
\\
$(\omega_{\pm})_{mn} = - g_{mp} \, (J_{\pm})^p{}_n$
&
$(\omega'_{\pm})_{mn} = - g'_{mp} \, (J'_{\pm})^p{}_n$
\\ \hdashline
$\hat{\nabla}_m g_{np} = 0$ 
&
$\hat{\nabla}'_m g'_{np} = 0$ 
\\
$\hat{\nabla}_m (J_{\pm})^p{}_n = 0 \Leftrightarrow \hat{\nabla}_m (\omega_{\pm})_{np} = 0$ 
&
$\hat{\nabla}'_m (J'_{\pm})^p{}_n = 0 \Leftrightarrow \hat{\nabla}'_m (\omega'_{\pm})_{np} = 0$
\\ \hdashline
$(N_{\pm})^p{}_{mn} = \mp (\Scr{C}_{\pm})^p{}_{mn}$ 
(vanish if $[\delta, \hat{\delta}] \Scr{L} = 0$)
&
$(N'_{\pm})^p{}_{mn} = \mp (\Scr{C}'_{\pm})^p{}_{mn}$ (vanish if $(N_{\pm})^p{}_{mn} = 0$)
\\ \hline
\end{tabular}
}
\end{center}
\end{table}

\end{appendix}

\phantomsection

\end{document}